%% file: main.tex
\documentclass[acmsmall,screen,nonacm]{acmart}
\ifdefined\pdfoutput\pdfoutput=1\fi
\AtBeginDocument{%
  }

\setcopyright{none}

\input{sections/00_prelude}

\begin{document}

\title{Hazel Prover: A Classroom Proof Assistant for Learning Structural Induction}
\titlenote{In submission to OOPSLA 2026.}

\author{Matthew Keenan}
\email{mckeenan@umich.edu}
\affiliation{%
  \institution{University of Michigan}
  \city{Ann Arbor}
  \country{USA}
}
\author{Nishant Kheterpal}
\email{nskh@umich.edu}
\affiliation{%
  \institution{University of Michigan}
  \city{Ann Arbor}
  \country{USA}
}
\author{Jean-Baptiste Jeannin}
\email{jeannin@umich.edu}
\affiliation{%
  \institution{University of Michigan}
  \city{Ann Arbor}
  \country{USA}
}
\author{Cyrus Omar}
\email{comar@umich.edu}
\affiliation{%
  \institution{University of Michigan}
  \city{Ann Arbor}
  \country{USA}
}

\renewcommand{\shortauthors}{Keenan et al.}

\begin{abstract}
    Proof assistants offer instant feedback and incremental proof scaffolding to users. Both of these features have long held promise in improving mathematics education in classroom settings, where manual grading is slow and costly, and students often struggle with knowing how to proceed at various points in their proof. In practice, proof assistants have been difficult to deploy effectively in classroom settings due to two main concerns: (i) students struggle to pick up the intricacies of full-scale proof assistants; and (ii) proof assistants provide ineffective assistance in support of student learning, as measured in terms of knowledge transfer to on-paper assessments without the tool.

    We present {\bf Hazel Prover}, a classroom proof assistant for teaching equational and inductive reasoning about programs with a design informed by a set of criteria encompassing ease-of-use of the tool, student engagement with the underlying mathematical ideas, knowledge transfer to pen-and-paper proof, and classroom logistics. We synthesized these criteria from observations made in prior deployments of proof assistants to the classroom.

    We engaged in an iterative design and evaluation process, deploying Hazel Prover in two different classes and conducting distinctively in-depth analyses of fine-grained usage logs, survey data, and student exam responses after each deployment. 
    Our analysis demonstrates that students were able to learn to use the tool effectively after a brief initial exploratory period, and that students became more capable with the mechanics of inductive proof as they progressed through problems. However, the first design iteration did not effectively achieve transfer to pen-and-paper proofs. We hypothesized from our analyses that this was due to the tool offering too much assistance to students in the equational reasoning steps. Based on this negative result, we enforced more manual student engagement with equational steps effective transfer in the second deployments.
    We believe that our analyses offer generalizable insights relevant to the designers of future classroom proof assistants for a variety of mathematical domains.
\end{abstract}

\maketitle

\input{sections/10_introduction}
\input{sections/25_design_goals}
\input{sections/30_by_example}
\input{sections/40_evaluation}
\input{sections/50_related_work}
\input{sections/60_conclusion}
\bibliographystyle{ACM-Reference-Format}
\bibliography{sections/99_references}

\end{document}

%% file: sections/00_prelude.tex
\newcommand{\hzlinline}[1]{\texttt{#1}}

\newtheorem{criterion}{Criterion}
\newcommand*{\ctnref}[1]{\hyperref[#1]{Criterion~ \autoref{#1} (\nameref{#1})}}

\usepackage{todonotes}
\setuptodonotes{inline}

\usepackage{tikz}
\usetikzlibrary{arrows.meta, fit, tikzmark}

\setcitestyle{authoryear}

\usepackage{subcaption}


%% file: sections/10_introduction.tex

\begin{figure}[p]
    \centering
    \includegraphics[scale=0.09]{"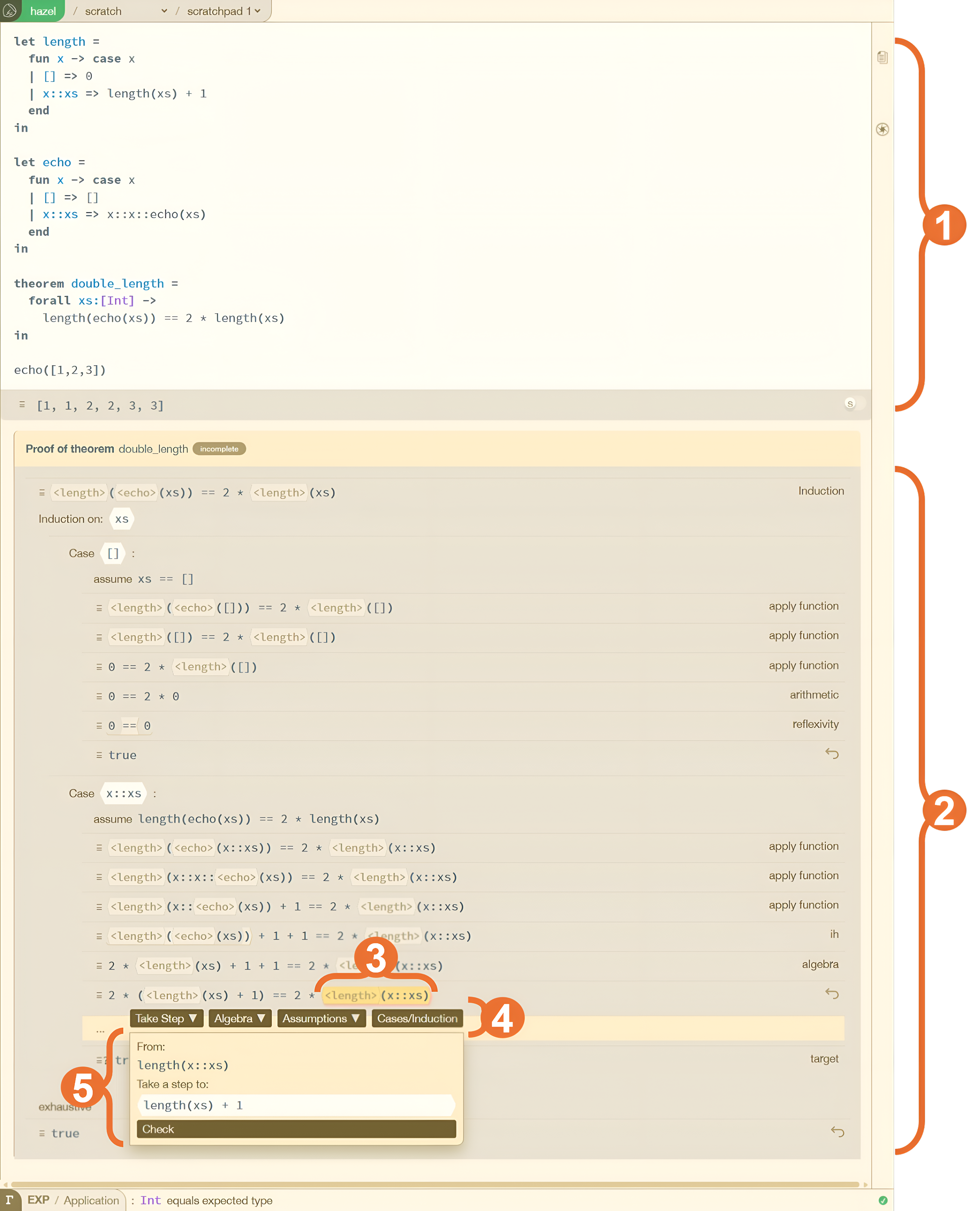"}
    \caption{An overview of the user interface of Hazel Prover showing: (1) an editor to write programs and theorems and see evaluation results, (2) space to prove theorems interactively, (3) a selection in the current goal of the theorem, (4) buttons representing possible next steps of the proof involving the given selection, and (5) the rewrite editor where students can propose rewrites of the current selection.}
    \label{fig:hero}
\end{figure}



\section{Introduction}

Learning to write detailed proofs is a central goal of many mathematics and computing courses. Traditionally, students write their proofs on paper and submit their attempts to graders for feedback. Paper proofs, however, impose several well-attested barriers to students developing a good mental model of what constitutes a correct proof. Graders are often inconsistent \cite{miller_how_2018, moore_mathematics_2016}, students rarely look at slow grader feedback unless explicitly asked to \cite{byrne_student_2018}, and while students can often correctly revise the proof in response to feedback they often fail to correctly explain why they needed to revise the proof \cite{byrne_student_2018, stylianides_research_2017}. This is all a far cry from the immediate, specific feedback a student receives from a compiler.

These observations have tempted many instructors to consider introducing proof assistants into the classroom. 
Full-scale proof assistants such as Lean~\cite{leanprover} and Rocq~(previously Coq)~\cite{coq_development_team_coq_2024} offer a much tighter feedback cycle, but in the classroom setting, they create extra barriers to students' writing the proofs, and then learning to prove independent of the tool. In particular, they impose a much higher syntactic and semantic burden, where students must become familiar with the linguistic intricacies of the proof assistant and its associated tooling~\cite{10.1145/3758317.3759679}. This creates a steep learning curve, and would require teachers to devote significant class time to teaching just the proof assistant itself. This would only be acceptable in classes where learning to use a theorem prover itself is a learning goal.

In addition, full-scale proof assistants in classrooms have been observed to leave students unable to transfer their knowledge to pen-and-paper, i.e. to produce the same reasoning steps, which is critical for exam-based assessment~\cite{knobelsdorf_theorem_2017, bohne_learning_2018}. 
The automation of certain tasks in full-scale proof assistants could be trivializing the reasoning that students need to spend time practicing.

Given these issues, there has been a long tradition of special-purpose proof assistants designed specifically for the classroom (with recent surveys by \citet{keenan_learner-centered_nodate} and \citet{minh_proof_2025}). In this paper, we focus on classroom proof assistants specialized for learning step-by-step equational and inductive reasoning about typed functional programs. We review two other tools with similar goals \cite{lodder_providing_2020, xu_emmy_nodate} and the broader space of classroom proof assistants, including for domains that we do not target such as natural deduction style derivations and real analysis, in \autoref{sec:related_work}. Although prior efforts have reported some success, as far as we are aware, prior work has not thoroughly investigated whether special-purpose classroom proof assistants like these suffer from the same transfer problems as full-scale proof assistants. 

In this paper we introduce Hazel Prover, or simply the Prover. It is implemented as an extension to Hazel, a live typed functional programming environment~\cite{omar_live_2019}. An overview of the Prover interface is given in \autoref{fig:hero}. We explain the design goals in detail in \autoref{sec:ctns} and the Prover implementation in detail in \autoref{sec:by_example}. In brief, students (or instructors) can interleave functional programs and theorem specifications inside Hazel's live functional programming environment (rather than relying on instructor-provided axiomatizations of function definitions in the two closely related tools). Specifications introduced with the \hzlinline{theorem} keyword appear at the bottom of the screen as proof obligations. Students interact with these proof obligations by highlighting sub-expressions and choosing whether to rewrite the expression or perform an induction. Rewrites can be evaluation steps (in the tradition of single-step evaluators like DrScheme \cite{clements2001modeling}), algebraic steps, or applications of equational assumptions or lemmas previously added to the context. As students write the proof, they are syntactically manipulating only the {\it program} syntax. There is no syntax specific to the proof assistant itself; all other proof logic is expressed via a graphical user interface. We claim that this design leads to a proof interface that not only requires minimal special training for students to pick up, but also closely follows the format and engagement pattern of proofs students write on paper in programming languages and formal methods courses. 

We engaged in an iterative design process by deploying Hazel Prover to two related programming languages courses in consecutive semesters, the first targeting upper-level undergraduates and the second early-stage graduate students (who had minimal experience with the ideas being taught). Both courses were taught by the same instructor, who taught the material relevant to this paper in substantially the same way. After each deployment, we conducted a distinctively thorough analysis of fine-grained usage logs, survey data, and exam data to understand the detailed time course of student learning, both of the tool itself and of the mathematical skills we hope students will master, both \emph{in situ} and when asked to transfer their knowledge to an on-paper exam (\autoref{sec:evaluation}).

After finding in our first deployment that both students' self-reported confidence in writing proofs without the tool, and their scores on the on-paper exams, were lower than we had hoped (and had observed in previous semesters where all instruction was on-paper), we iterated on the design. In particular, the biggest change was that in the second iteration, students had to explicitly write out evaluation steps instead of simply clicking possibly reducible sub-expressions. This increase in the required engagement with the tool substantially improved transfer, leaving students able to complete pen-and-paper proofs on exams as effectively as in previous semesters, despite having practiced the ideas exclusively using Hazel Prover without the need for costly manual grading. 

This paper makes both specific and generalizable contributions to the field of classroom proof assistants. The specific contributions are the design, implementation, and evaluation of Hazel Prover, the first classroom theorem prover that has been empirically established to enable student learning of step-by-step equational and inductive reasoning about functional programs in a manner that has been demonstrated through classroom deployments to be quick to learn, to support student learning of the underlying formal skills over the span of a few practice problems, and to support student knowledge transfer to on-paper proofs like those that would appear on exams. Hazel Prover itself is likely to be of interest to instructors from the programming languages and formal methods research communities, and it has been designed to support practical deployments today.

The generalizable contributions include a set of learner-centered design goals that can help guide the design of classroom theorem provers in other domains, a demonstration of how fine-grained log analysis, e.g. of backtracking behavior, can help instructors understand the time course of student learning of individual skills relevant to individual milestones within a proof, and non-obvious lessons about the extent to which even small amounts of seemingly tedium-saving automation can lead to dramatic transfer failures. We conclude with a discussion of these and other lessons from our study, and potential directions for future work.

%% file: sections/25_design_goals.tex
\section{Design Goals}
\label{sec:ctns}

We identified two broad key goals to ensure that any classroom proof assistant is useful as a pedagogical tool: \textbf{learning}: the tool should help students refine and improve their mathematical proof skills; and \textbf{usability}: the tool should not create unnecessary frictions in writing proofs. In order to guide our design, we have expanded upon each goal with criteria informed by previous classroom proof assistants, refining a previous workshop survey \cite{keenan_learner-centered_nodate}. We are guided by these design criteria in our design, and we believe that these criteria are transferrable to guide design of classroom proof assistants for other domains.


\subsection{Learning}
\label{sec:dg_learning}

Learning is definitionally the main goal of a classroom proof assistant. When assessing learning, it is important to do so in the context of one's specific learning objectives. One particularly salient decision that must be made is whether the objective is to teach users to write proofs using proof assistants, or to teach users to prove independently of proof assistants. In this Section, we give five design criteria for learning. The first three are core criteria we believe to be applicable regardless of whether one is trying to teach students to prove without the tool. The final two criteria are specifically focused on the transfer of students' learning from the tool to pen-and-paper.

\subsubsection{Learning with the Tool}

We will begin with the learning criteria that relate to learning within the tool. Previous deployments of proof assistants to the classroom have observed that the restricted search space prevents students from giving up \cite{kerjean_utilisation_2022}.

\begin{criterion}
    [Supporting Active Exploration]
    \label{ctn:discover}
    Classroom proof assistants should provide affordances that allow students to actively explore various potentially productive proof ideas and avoid leaving students unsure for an extended period of time about how to explore in another direction.
\end{criterion}

Another helpful aspect of proof assistants is that they create a more predictable model for grading while the student is learning. It has been shown that professors teaching proof do not always effectively communicate their values and expectations for proofs to students \cite{dawkins2017values, e0c7fdf9-b8a2-3d03-96c4-fbf15b9d2be0}. A proof assistant's expectations for a proof can be a lot more transparent: \citet{avigad_learning_2019} observed with his students that ``at least with Lean they knew what the rules were, whereas they could not anticipate how their informal proofs would be received by the person grading them''.


\begin{criterion}
    [Consistent model]
    \label{ctn:predictable}
    As students use the tool, it should provide both positive and negative feedback that allows students to build a consistent mental model of the mathematical rules of the domain. This feedback should be provided for partial proofs, not just once a proof is complete.
\end{criterion}

Another key prerequisite is that students must be engaged in order to learn. In one deployment of a bespoke classroom proof assistant, 21\% of their students reported being able to complete some exercises by random clicking, without understanding anything~\cite{kerjean_utilisation_2022}. If students are able to complete exercises in such a way, learning is ineffective. The classroom proof assistant should therefore require students to think carefully about the mathematical ideas of the domain. This goal requires care when considering the value of automation or automatic hint systems.

\begin{criterion}
    [Require Engagement]
    \label{ctn:engagement}
    The assistant must require genuine cognitive engagement with the mathematical ideas in order to complete exercises.
\end{criterion}


\subsubsection{Transfer to Pen-and-Paper}
In many (but not all) classroom settings, students are expected to be able to write proofs on-paper. This is both to support exam-based assessment and, for many instructors, it serves as a proxy for whether the student has truly internalized the ideas.

Computer science education literature uses the terms {\it scaffold} and {\it support} to describe the distinction between tools designed for transfer and tools not designed for transfer. Where a scaffold is a support that ``facilitates the student learning to achieve the goal or action without the support in the future'' \cite{guzdial_software-realized_1994}. Prior work has shown that often proof assistants that are effective supports can still fail to prepare students to write proofs without the tool~\cite{knobelsdorf_theorem_2017,pierce_lambda_2009}.

Transfer has been observed to be a problem in previous deployments of proof assistants to the classroom \cite{knobelsdorf_theorem_2017, bohne_learning_2018}. We ourselves also observed this difficulty particularly in our first deployment.

One blocker to transfer is the distance in appearance and structure between proof assistant proofs and pen-and-paper proofs. For example it has been shown to require a lengthy process to map students' knowledge of Rocq proofs onto pen-and-paper proofs \cite{bohne_learning_2018}. To aid transfer, we instead aim to lay proofs out in a manner recognizable to a proof class.

\begin{criterion}
    [Notation Similar to Written Proofs]
    \label{ctn:notation}
    Proofs in the tool should be laid out similarly to how students would be expected to produce proofs.
\end{criterion}

Until reviewing our first deployment we did not have the next criterion. In our first deployment we observed students were having difficulty writing proofs on pen-and-paper after learning using Hazel Prover. This observation, coupled with the fact that another tool we were using that semester, Hazel Deriver \cite{zhong_hazel_2025}, scored better on transfer, motivated us to add an extra criterion.

We want to specify that a classroom proof assistant must not only force the user to engage, but the engagement pattern must be similar enough to pen-and-paper. This criterion motivated a major iteration of the tool between the two deployments. In our first deployment, we noticed students weren't sufficiently engaged in writing the proof, in particular they weren't sufficiently engaged in writing out the evaluation steps as part of a proof. The design change itself is described in \autoref{sec:evstep}.


\begin{criterion}
    [Engagement Similar to Written Proofs]
    \label{ctn:engagement2}
    Writing a proof in the tool should force users to think through all the steps they would need to learn to think through in order to write a proof on paper.
\end{criterion}


\subsection{Usability}
\label{sec:dg_usability}

Our second key objective is usability. In the classroom context teachers want to waste as little time as possible teaching students the specifics of the tool and its installation, so that students can spend as much time as possible engaging with the content.

While earlier attempts at classroom deployments of proof assistants often had to deal with the difficulties of explaining to students how to set up specific command line tooling for proof assistants \cite{wemmenhove_waterproof_2024}, this setup cost has become significantly lower as tools have become easier to install in recent years. Now even full-scale proof assistants can be embedded in websites \cite{gallego_arias_jscoq_2017} that require no installation. We have also taken this approach, providing a fully web-based editor.

\begin{criterion}
    [Easy Setup]
    \label{ctn:setup}
    Installation and setup of the tool should be as straightforward as possible.
\end{criterion}

Since teachers want to focus on teaching content, and not teaching a new tool, classroom proof assistants must be approachable to students (and to instructors in many cases) with no programming background after minimal training---perhaps no more than a portion of a 50-minute lab or discussion.

\begin{criterion}
    [Minimal Training]
    \label{ctn:curve}
    Classroom proof assistants should require minimal training specific to the tool.
\end{criterion}

Proof formalization can be a tedious task, especially since the goal is to be rigorous, this can often involve justifying things that you would not normally need to justify on paper, or can create large numbers of tedious cases to write out. For a proof assistant to be usable in the classroom, it must not create extra work that does not help with the class's learning objectives. We must, however, be careful to balance this desire to minimize tedious tasks against \ctnref{ctn:engagement}, as sometimes forcing students to spend more time on tasks relevant to the learning objectives is good for their learning.

\begin{criterion}
    [Minimal Tedium]
    \label{ctn:tedium}
    Classroom proof assistants should not obligate proof
steps that would be elided on paper, or require students to perform didactically trivial but long tasks such as copying out lines of text.
\end{criterion}

%% file: sections/30_by_example.tex
\section{Hazel Prover}
\label{sec:by_example}

We will now walk through all major features of the Hazel Prover, using the \hzlinline{echo} function as our running example. We will show the features of both deployments, including the iteration we made. This example was used in a live demo during a lecture in the first class deployment to teach induction over lists. An overview of the full user interface is given in \autoref{fig:hero}.

\subsection{Function Definitions in Hazel}
We begin by defining the \hzlinline{length} and \hzlinline{echo} functions using the Hazel editor. In classes where we teach functional programming, students will be familiar with the Hazel editor for programming exercises. We hope this adds to the authenticity of the exercise as well -- students can see that these functions are defined the same way they would write code. The \hzlinline{echo} function duplicates each element in a list and returns a list double the length of the original.



\begin{center}
    \includegraphics[scale=0.45]{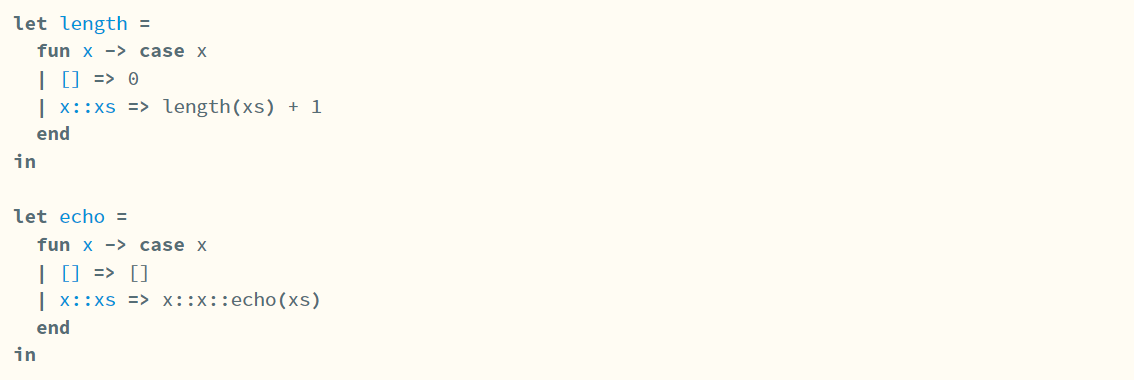}
\end{center}

Both functions are defined recursively. The main body of the function uses \hzlinline{case} to pattern match on the input to the function, \hzlinline{xs}. Cases are separated by pipes, \hzlinline{|}. Each case consists of a pattern and a corresponding branch. The \hzlinline{[]} pattern matches empty lists, and \hzlinline{x::xs} (read ``x cons xs'') matches non-empty lists, which consist of a head element, \hzlinline{x}, and a tail list, \hzlinline{xs}.

In order to set up proof exercises, these functions can be locked so that students are unable to edit them. Scratch space with live evaluation is provided so that students can try out functions. In our second deployment we also included one exercise where students need to write their own implementation of a function before proving the correctness of their implementation.

\subsection{Proof Specification}

Specification is also done inside the Hazel editor,
%
%
%
%
Which we have extended with the \hzlinline{theorem} and \hzlinline{forall} constructs. For example, the following theorem, named \hzlinline{echo\_length}, states that \hzlinline{echo} doubles the length of its input:


\begin{center}
    \includegraphics[scale=0.45]{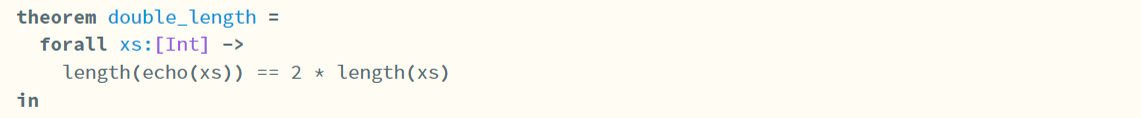}
\end{center}

The \hzlinline{theorem} construct takes a theorem name and a boolean expression extended with quantifiers.


We will pause here and discuss our decision to make the theorems use boolean expressions. A perhaps more conventional option would be to use types to specify theorems. The Curry-Howard correspondence \cite{wadler2015propositions} between types and propositions is a well-known result that gives a correspondence between common logical connectives and connectives in types. Using this system, ``and'' would be encoded as products (tuple types), ``or'' would be encoded as sums (data constructors) ``implies'' would be encoded as arrows (function types), and we would need to add a new construct to Hazel's types to represent ``forall''.

The advantages of this system are the elegance of this theory and the fact that we can reuse our expression editing capabilities to support proof editing. There are, however, some disadvantages:

\begin{itemize}
    \item 
    It could expose students to the Curry-Howard correspondence just as they begin writing proofs. This would steepen the learning curve, something we are explicitly trying to avoid in \ctnref{ctn:curve}. 

    \item
    It would involve a large development effort to add dependent types to Hazel. Once types can involve expressions, we need to add evaluation logic into the type-checking logic. We would also need to develop new theories to support Hazel's holes in dependent types.
\end{itemize}

Instead we decided to keep specifications separate from the type system, using a separate \hzlinline{theorem} construct that uses boolean expressions, extended with a \hzlinline{forall} construct. (This construct is only intended to be used in specification, and not in programs, so it has no evaluation semantics.) 




\subsection{Proof Steppers}

For every theorem construct at the top-level of the hazel program, a proof stepper is opened at the bottom of the editor. All variables bound in \hzlinline{forall} quantifiers are introduced to the scope of the proof automatically, to match the way introduction would be elided in a written proof for this class.

\begin{center}
    \includegraphics[scale=0.45]{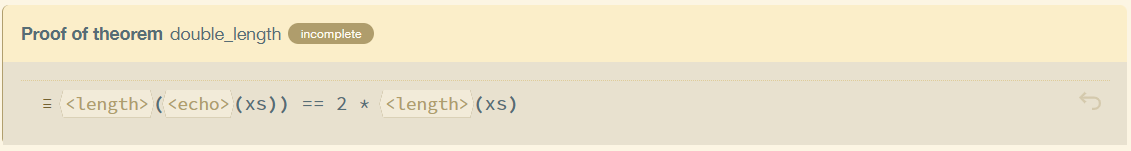}
\end{center}

The proof stepper shows an expression, and if the student can successfully step this boolean expression to \hzlinline{true}, the theorem is marked as complete. Students can directly interact with this expression by clicking and dragging to highlight subexpressions. Upon highlighting a subexpression, buttons are shown below the expression to show the user's options at this point in the proof.

\begin{center}
    \includegraphics[scale=0.45]{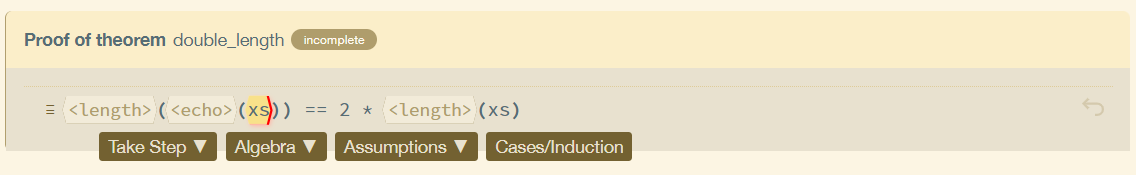}
\end{center}

\subsection{Induction}

In order to begin a proof by induction, the student can highlight \hzlinline{xs} and click the ``Cases/Induction" button. This will open up an induction on the possible values of \hzlinline{xs}.

\begin{center}
    \includegraphics[scale=0.45]{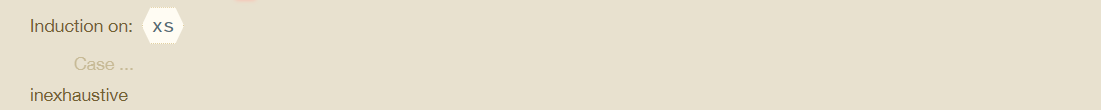}
\end{center}

Because learning how to format an induction is an explicit goal of the class we are teaching, we made the decision to require students to add all the cases to the induction manually in the \hzlinline{Case}~\hzlinline{...} field. As per \ctnref{ctn:engagement}, this forces students to engage more with the structure of an inductive proof. This also allows students to use arbitrary pattern matching in their case split, so they can, for example use \hzlinline{[]}, \hzlinline{[x]}, \hzlinline{x::y::zs} instead of just \hzlinline{[]} and \hzlinline{x::ys}. Feedback at the bottom of the case split tells the student whether their split is exhaustive.

\begin{center}
    \includegraphics[scale=0.45]{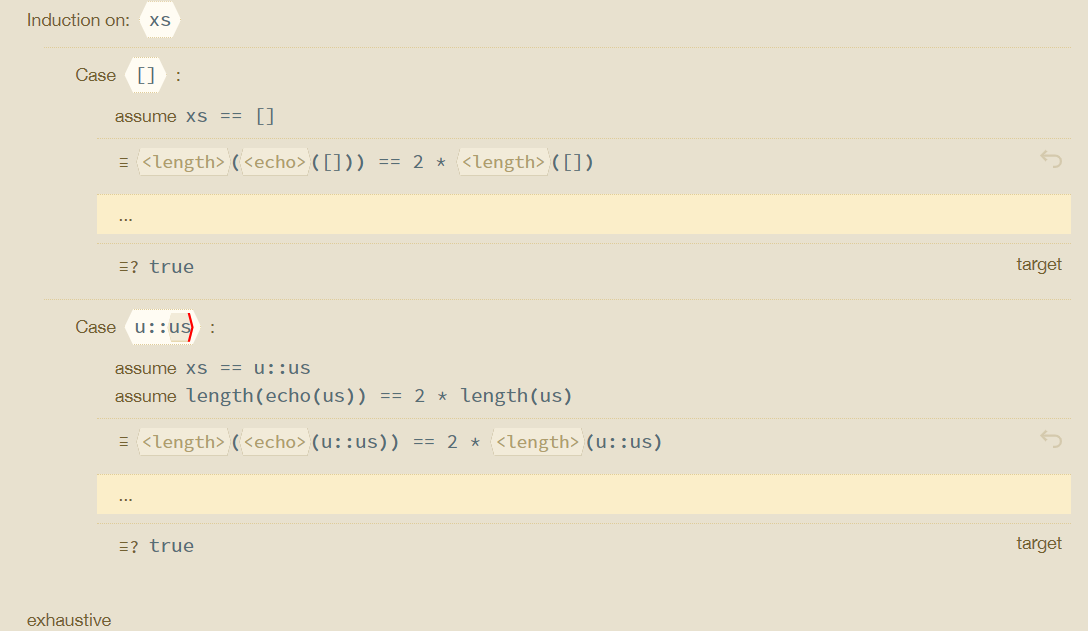}
\end{center}

An induction step steps any expression to \hzlinline{true}, but creates a proof obligation for every case of the induction. In each of these obligations you must show that the original expression evaluates to \hzlinline{true}. Within each induction step you have access to the match equation, and an inductive hypothesis for every sub-expression of the pattern with the same type as the original pattern. In this case, since the variable \hzlinline{us} in the \hzlinline{u::us} case also has type \hzlinline{[Int]}, we are given one inductive hypothesis to work with. These hypotheses are displayed to the user just above each proof obligation.

\subsection{Evaluation Steps}
\label{sec:evstep}

Once inside the induction, the student needs to solve each of the proof obligations using equational reasoning.
A significant proportion of this equational reasoning will be tracing the evaluation of the function. In many full-scale proof assistants, these steps are fully automated and hidden from the user as part of a normalization routine \cite{paulinmohring:hal-01094195}, but since learning how this evaluation works is an explicit learning goal of our class, we instead explicitly write out the evaluation steps.

For the first deployment we used Hazel's {\it single stepper}, which highlights syntax that can take an evaluation step green. When the user double-clicks on the green syntax, it takes a single evaluation step. By highlighting all the steps that can be taken, we are making it clear to the user the full space of evaluation they can explore (\ctnref{ctn:discover}). In the base case of our example, we can trace this obligation to its conclusion by repeated clicking:

\begin{center}
    \includegraphics[scale=0.45]{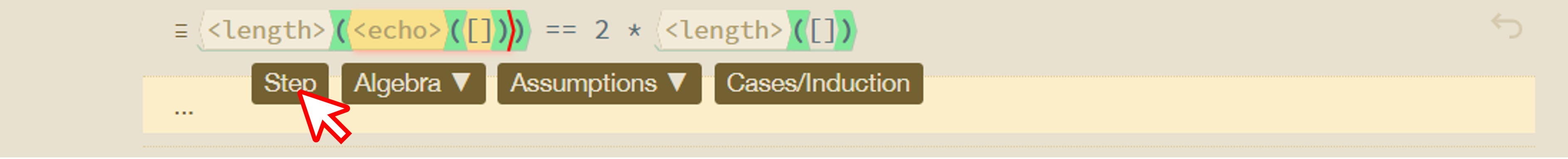}
\end{center}

Hazel's single stepper originally disallowed stepping function applications in locations where the argument hasn't finished evaluation (such as the application of length function on the LHS above), but we have removed this restriction to permit general symbolic reasoning. 


In the first deployment, we found students' transfer to pen-and-paper to be lower than we had hoped. We hypothesized that this could mean double-clicking highlighted expressions didn't adequately meet \ctnref{ctn:engagement}. For our second deployment, we instead required that students highlight a sub-expression and write out what that expression steps to, they can then press a ``check'' button that will check whether this is a valid step. If it is allowed, it will be inserted into the proof. Results show this appears to have improved students' engagement \autoref{sec:survey_transfer}, and their pen-and-paper exam results \autoref{sec:exam_results}.

\begin{center}
    \includegraphics[scale=0.45]{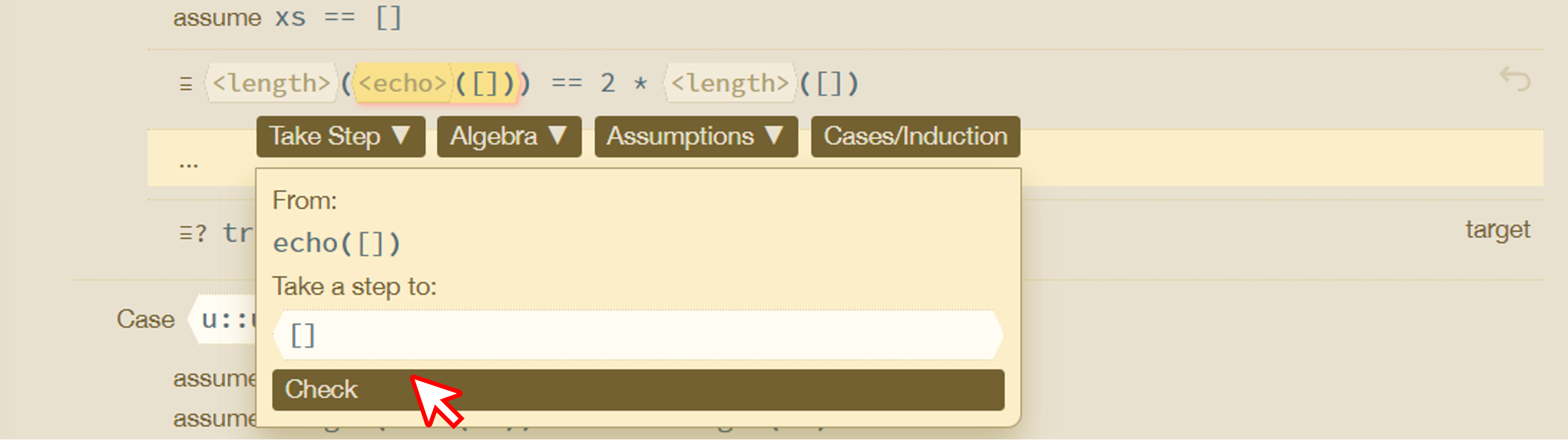}
\end{center}


\subsection{Algebra Steps}
\label{sec:algebrite}

Before we can use the inductive hypothesis, we need to get our current expression in the right form, which requires an algebra step.

\begin{center}
    \includegraphics[scale=0.45]{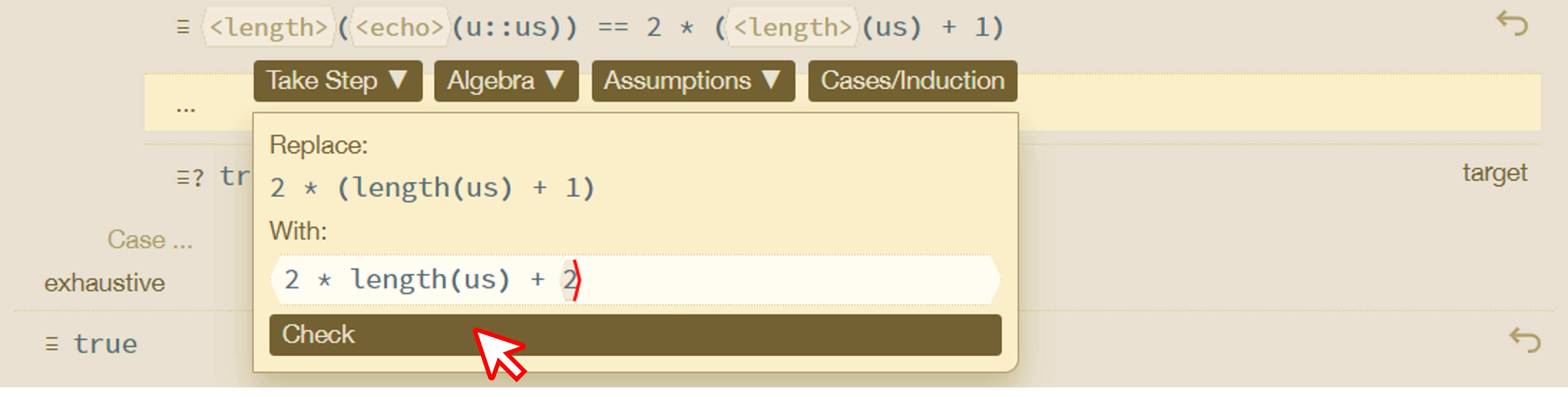}
\end{center}

We need algebra steps because, as above, proof goals with symbolic terms cannot always be fully evaluated. Since we are not explicitly aiming to teach students basic algebra (we assume they are already proficient), we allow students to fill in arbitrary algebraic rewrites, and use an external computer algebra checker (Algebrite \cite{noauthor_algebrite_nodate}) to check their work. This removes the tedium and often unnecessary complexity of proving these algebraic rewrites from first principles. By only giving the solver access to the arithmetic and arithmetic operators, and leaving Hazel's control flow and list primitives opaque, we ensure that students cannot simply rewrite the entire proposition to true and have the solver check it for them (in line with \ctnref{ctn:engagement}).

\subsection{Assumption Steps}

Now that the expression is in the right form, we can then find the inductive hypothesis from the assumptions box. After this, all we need is a few more evaluation steps and we will eventually also reach \hzlinline{true}, completing the proof.

\begin{center}
    \includegraphics[scale=0.45]{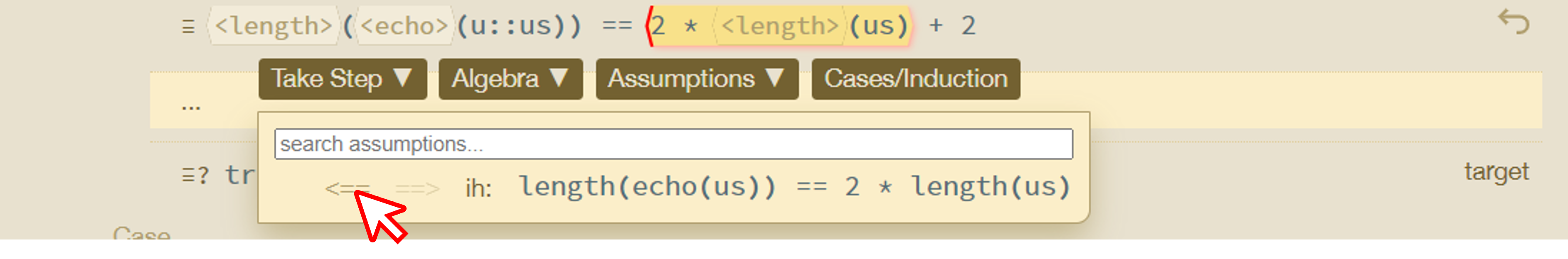}
\end{center}

The assumption box includes assumptions from the current context (including previously proven theorems, and inductive hypotheses).
When the box is first opened, only assumptions where one side of the equality matches the selected expression are shown. The student can however also search for the name of an assumption to see the other assumptions that don't apply.


%% file: sections/40_evaluation.tex
\section{Classroom Evaluation}
\label{sec:evaluation}

Our research questions are derived from our design goals given in \autoref{sec:ctns}. We are interested in: 

\begin{quote}
{\bf RQ1a (Learning):} Can we observe students learning equational reasoning and induction-related skills over time while using Hazel Prover?
\end{quote}

\begin{quote}
{\bf RQ1b (Transfer):} Are students able to transfer the skills they have learned by using Hazel Prover to on-paper exams?
\end{quote}

\begin{quote}
{\bf RQ2 (Usability)} Are students able to quickly learn to use Hazel Prover's user interface?
\end{quote}

We evaluated Hazel Prover in two authentic classroom deployments, where students used the tool to complete homework assignments. From this we obtained three kinds of data: log data (\autoref{sec:action_log}), survey responses (\autoref{sec:survey}), and exam results (\autoref{sec:exam_results}). We have summarized how these results answer each of our research questions in \autoref{sec:summary_of_findings}.


\subsection{Classes and Learning Objectives}
\label{sec:class}

We evaluated Hazel Prover in two different classes in two consecutive semesters. The first was an upper-level undergraduate class, and the second was a graduate class. We used the results of the first deployment to inform a design iteration (the most significant changes are described in \autoref{sec:evaluation}) before the second deployment. Care needs to be taken, however, when we make comparisons between the two classes, because they were 1) operating at different levels, 2) had different levels of familiarity with Hazel before using the prover, and 3) had different levels of instruction.

The full specification for all exercises in both classes is given in the supplemental material.

\subsubsection{First Deployment}

We first evaluated Hazel Prover in the Fall 2025 offering of an upper-level undergraduate programming language class. The class aims to give students a foundation in programming language design, programming language theory, language prototyping, and formal reasoning about program behavior. As a prerequisite to these topics, the class also has to teach the basics of functional programming and structural induction. There were 25 students in the class that semester.

Hazel \cite{omar_live_2019} was used for the first six assignments of the class. Assignment 1 only had coding exercises, assignment 2 had one coding and two Hazel Prover induction exercises, and assignments 3-6 had Hazel Deriver exercises (we compare Hazel Prover and Hazel Deriver below).

The first Prover exercise, which we will refer to as Ex1, defines a \hzlinline{compose} function over two functions and a \hzlinline{map} function that applies a function to every element in a list. Students were asked to prove that mapping two different functions over a list returned the same result as mapping a composition of the two functions over the list.

The second Prover exercise defined a \hzlinline{rev} function that reverses a list, defined in terms of a \hzlinline{snoc} (backwards cons) function that adds an element to the end of a list. Students were asked to show that reversing a list twice returns the original list. Note that this requires first proving a lemma relating \hzlinline{rev} and \hzlinline{snoc} by induction, which we will call Ex2a, and then using that lemma in another induction, which we will call Ex2b. The specification for the required lemma was given to students, but they had to prove both inductions.

\subsubsection{Second Deployment}

We then evaluated our iteration of Hazel Prover in the Winter 2026 offering of a graduate level programming languages class. There were 16 students in the class that semester.

Hazel Deriver was used in the first assignment of this class, and Hazel Prover and Hazel Deriver were both used in the second assignment. Hazel Deriver is separated from Hazel's functional programming language, so unlike in the undergraduate class, students in the graduate class did not have experience writing functional programs in Hazel's editor before completing this exercise.

The second assignment for this class has the same two proof exercises as the previous, with one additional proof exercise. 

This proof exercise defined Peano numbers using ADTs, along with a \hzlinline{convert} function that takes a Peano number and returns Hazel's integer type. Students were given an addition function, and a proof that the addition function was correct using the \hzlinline{convert} function. They were then asked to give their own implementation of Peano multiplication, and prove that it was correct. Unlike the previous two problems, this problem requires students to not only write a proof, but also to write their own implementation.



\subsection{Student Interaction with the tool}
\label{sec:action_log}

\definecolor{boxblue}{HTML}{5B8DB8}
\definecolor{boxrust}{HTML}{7A2E10}
\definecolor{boxteal}{HTML}{1A5E6B}
\definecolor{boxpeach}{HTML}{C07000}
\definecolor{boxolive}{HTML}{9BBB59}
\definecolor{boxgreen}{HTML}{3A8E8A}

\begin{table*}[t]
\centering
\small
\renewcommand{\arraystretch}{1.8}
\begin{tabular}{l@{\hskip 12pt}l@{\hskip 12pt}|c@{\hskip 10pt}c@{\hskip 10pt}c@{\hskip 10pt}c|c@{\hskip 10pt}c@{\hskip 10pt}c@{\hskip 10pt}c}
 &  & \multicolumn{4}{c|}{First ($n=18$)} & \multicolumn{4}{c}{Second ($n=14$)} \\
Backtracking & Ex. & Out. & Base & Pre-IH & Post-IH & Out. & Base & Pre-IH & Post-IH \\
\hline
Any backtracking & 1 & \tikzmarknode{FA1o}{100} & \tikzmarknode{FA1b}{39} & \tikzmarknode{FA1p}{94} & \tikzmarknode{FA1q}{44} & \tikzmarknode{SA1o}{79} & \tikzmarknode{SA1b}{21} & \tikzmarknode{SA1p}{36} & \tikzmarknode{SA1q}{14} \\
 & 2a & \tikzmarknode{FA2ao}{72} & \tikzmarknode{FA2ab}{50} & \tikzmarknode{FA2ap}{89} & \tikzmarknode{FA2aq}{56} & \tikzmarknode{SA2ao}{43} & \tikzmarknode{SA2ab}{7} & \tikzmarknode{SA2ap}{7} & \tikzmarknode{SA2aq}{0} \\
 & 2b & \tikzmarknode{FA2bo}{50} & \tikzmarknode{FA2bb}{0} & \tikzmarknode{FA2bp}{61} & \tikzmarknode{FA2bq}{6} & \tikzmarknode{SA2bo}{7} & \tikzmarknode{SA2bb}{0} & \tikzmarknode{SA2bp}{7} & \tikzmarknode{SA2bq}{0} \\
 & 3 & \tikzmarknode{FA3o}{---} & \tikzmarknode{FA3b}{---} & \tikzmarknode{FA3p}{---} & \tikzmarknode{FA3q}{---} & \tikzmarknode{SA3o}{21} & \tikzmarknode{SA3b}{0} & \tikzmarknode{SA3p}{21} & \tikzmarknode{SA3q}{14} \\
\hline
$\hookrightarrow$ Rm induction & 1 & \tikzmarknode{FI1o}{89} & \tikzmarknode{FI1b}{17} & \tikzmarknode{FI1p}{44} & \tikzmarknode{FI1q}{17} & \tikzmarknode{SI1o}{50} & \tikzmarknode{SI1b}{14} & \tikzmarknode{SI1p}{29} & \tikzmarknode{SI1q}{7} \\
 & 2a & \tikzmarknode{FI2ao}{39} & \tikzmarknode{FI2ab}{0} & \tikzmarknode{FI2ap}{28} & \tikzmarknode{FI2aq}{0} & \tikzmarknode{SI2ao}{43} & \tikzmarknode{SI2ab}{7} & \tikzmarknode{SI2ap}{0} & \tikzmarknode{SI2aq}{0} \\
 & 2b & \tikzmarknode{FI2bo}{28} & \tikzmarknode{FI2bb}{0} & \tikzmarknode{FI2bp}{0} & \tikzmarknode{FI2bq}{0} & \tikzmarknode{SI2bo}{7} & \tikzmarknode{SI2bb}{0} & \tikzmarknode{SI2bp}{7} & \tikzmarknode{SI2bq}{0} \\
 & 3 & \tikzmarknode{FI3o}{---} & \tikzmarknode{FI3b}{---} & \tikzmarknode{FI3p}{---} & \tikzmarknode{FI3q}{---} & \tikzmarknode{SI3o}{14} & \tikzmarknode{SI3b}{0} & \tikzmarknode{SI3p}{21} & \tikzmarknode{SI3q}{0} \\
\hline
$\hookrightarrow$ Rm step & 1 & \tikzmarknode{FR1o}{72} & \tikzmarknode{FR1b}{33} & \tikzmarknode{FR1p}{78} & \tikzmarknode{FR1q}{39} & \tikzmarknode{SR1o}{0} & \tikzmarknode{SR1b}{0} & \tikzmarknode{SR1p}{0} & \tikzmarknode{SR1q}{0} \\
 & 2a & \tikzmarknode{FR2ao}{61} & \tikzmarknode{FR2ab}{50} & \tikzmarknode{FR2ap}{89} & \tikzmarknode{FR2aq}{56} & \tikzmarknode{SR2ao}{0} & \tikzmarknode{SR2ab}{0} & \tikzmarknode{SR2ap}{0} & \tikzmarknode{SR2aq}{0} \\
 & 2b & \tikzmarknode{FR2bo}{39} & \tikzmarknode{FR2bb}{0} & \tikzmarknode{FR2bp}{61} & \tikzmarknode{FR2bq}{0} & \tikzmarknode{SR2bo}{0} & \tikzmarknode{SR2bb}{0} & \tikzmarknode{SR2bp}{0} & \tikzmarknode{SR2bq}{0} \\
 & 3 & \tikzmarknode{FR3o}{---} & \tikzmarknode{FR3b}{---} & \tikzmarknode{FR3p}{---} & \tikzmarknode{FR3q}{---} & \tikzmarknode{SR3o}{0} & \tikzmarknode{SR3b}{0} & \tikzmarknode{SR3p}{0} & \tikzmarknode{SR3q}{0} \\
\end{tabular}
\renewcommand{\arraystretch}{1.0}

\vspace{2pt}
\begin{tikzpicture}[remember picture, overlay,
  box/.style={draw, rounded corners=2pt, inner sep=4pt, line width=0.7pt},
  boxwide/.style={draw, rounded corners=2pt, inner sep=7pt, line width=0.7pt},
  lbl/.style={font=\scriptsize\bfseries, anchor=south west},
]
  \node[box, draw=boxblue, fit=(FA1o)(FA2ao)(FA2bo)(FA3o)] (boxa1) {};
  \node[lbl, boxblue, anchor=south west] at (boxa1.south east) {\textsf{(a)}};
  \draw[-{Stealth[length=4pt]}, boxblue, thick] ([xshift=3pt, yshift=-3pt]boxa1.north west) -- ([xshift=3pt, yshift=3pt]boxa1.south west);

  \node[box, draw=boxblue, fit=(SA1o)(SA2ao)(SA2bo)(SA3o)] (boxa2) {};
  \node[lbl, boxblue, anchor=south west] at (boxa2.south east) {\textsf{(a)}};
  \draw[-{Stealth[length=4pt]}, boxblue, thick] ([xshift=3pt, yshift=-3pt]boxa2.north west) -- ([xshift=3pt, yshift=3pt]boxa2.south west);

  \node[box, draw=boxrust, fit=(FR1o)] (boxb) {};
  \node[lbl, boxrust, anchor=north east] at ([xshift=8pt]boxb.south east) {\textsf{(b)}};

  \node[box, draw=boxteal, fit=(FR1b)(FR2ab)(FR2bb)] (boxc1) {};
  \node[lbl, boxteal, anchor=north east] at (boxc1.south east) {\textsf{(c)}};
  \draw[-{Stealth[length=4pt]}, boxteal, thick] ([xshift=3pt, yshift=-3pt]boxc1.north west) -- ([xshift=3pt, yshift=3pt]boxc1.south west);

  \node[box, draw=boxteal, fit=(FR1q)(FR2aq)(FR2bq)] (boxc2) {};
  \node[lbl, boxteal, anchor=north east] at (boxc2.south east) {\textsf{(c)}};
  \draw[-{Stealth[length=4pt]}, boxteal, thick] ([xshift=3pt, yshift=-3pt]boxc2.north west) -- ([xshift=3pt, yshift=3pt]boxc2.south west);

  \node[box, draw=boxpeach, fit=(SR1o)(SR1q)(SR3o)(SR3q)] (boxd) {};
  \node[lbl, boxpeach, anchor=south west] at (boxd.south east) {\textsf{(d)}};

  \node[boxwide, draw=boxolive, fit=(FR1p)(FR1q)] (boxe) {};
  \node[lbl, boxolive, anchor=north] at (boxe.south) {\textsf{(e)}};
  \draw[-{Stealth[length=4pt]}, boxolive, thick] ([xshift=3pt, yshift=3pt]boxe.south west) -- ([xshift=-3pt, yshift=3pt]boxe.south east);

  \node[box, draw=boxgreen, fit=(FI1p)(FI2ap)(FI2bp)] (boxf) {};
  \node[lbl, boxgreen, anchor=south west] at (boxf.south east) {\textsf{(f)}};
  \draw[-{Stealth[length=4pt]}, boxgreen, thick] ([xshift=3pt, yshift=-3pt]boxf.north west) -- ([xshift=3pt, yshift=3pt]boxf.south west);

  \node[box, draw=boxgreen, fit=(SI1p)(SI2ap)(SI2bp)(SI3p)] (boxg) {};
  \node[lbl, boxgreen, anchor=south west] at (boxg.south east) {\textsf{(g)}};
  \draw[-{Stealth[length=4pt]}, boxgreen, thick] ([xshift=3pt, yshift=-3pt]boxg.north west) -- ([xshift=3pt, yshift=3pt]boxg.south west);
\end{tikzpicture}
\caption{Percentage of students performing backtracking actions in each part of each proof. ``Out.'' refers to actions taken outside the induction's cases, including actions taken before the induction was created, ``Base'' refers to the base case, ``Pre-IH'' refers to actions taken in the inductive step but before the inductive hypothesis was used, ``Post-IH'' refers to actions taken in the inductive step after the inductive hypothesis. Annotated regions (a)--(g) are referenced in the text.}
\label{tab:backtracking}
\end{table*}

\newcommand{\btref}[1]{\autoref{tab:backtracking}(#1)}


In order to evaluate students' learning (\autoref{sec:dg_learning}), we collected log data for both deployments. We could then replay these logs to observe how students completed the exercises. We performed a fine-grained analysis, splitting the proof up into a series of milestones, and categorizing actions, so we can get specific clues about which parts of the proof were most difficult; supporting this by watching some replays to see exactly what the students were struggling with. In particular we observed the proportion of students who performed backtracking action in various stages of the proof, this data is given in full in \autoref{tab:backtracking}. We observed a big change in the students' patterns of behavior between the two deployments, and we also take note of some usability issues that could be fixed in future deployments.

\subsubsection{Beginning the Induction}
\label{sec:log_begin_ind}

Students' logs for starting the induction appear to show that there were a few initial barriers to usability when students first loaded the page, but that these went down as the student used the tool more. This is evidenced in \btref{a} by a marked decrease in students' backtracking before setting up the correct induction. In the first deployment these went from 94\% of students backtracking the first exercise to 50\% of students backtracking in the last exercise. In the second deployment, it went from 33\% to 0\%. This large decrease can likely largely be explained by some discoverability issues students had at the start. We believe that this shows that there are some simple interventions we can make to improve usability, and in particular simplify the initial learning curve.

In order to start an induction, students needed to first select a part of the proof goal where they wish to perform the induction and then click the ``cases/induction'' button. In the first deployment, selection was a much less clear affordance than the green highlights which showed possible evaluation steps. This led to 72\% (\btref{b}) of these students taking an evaluation step before realizing it was unproductive and removing it. In the second deployment, without the green highlights, students were instead observed typing in the scratch space, as that was then the most obvious affordance. Some users in the second deployment specifically commented that it was ``not at all intuitive that you manipulated things by highlighting and then adding clauses". In future design iterations, we should look into making selection a clearer affordance, including possibly some text under incomplete goals instructing students that they can highlight part of the expression to proceed.

Another common confusion was over what needed to be selected to perform an induction. When you choose to do a proof by induction in Hazel Prover, it will auto-populate the induction variable with whatever was selected. 83\% of students in the first deployment and 60\% of students in the second deployment selected the entire expression in the induction at some point, and this created a confusing situation where you would see ``induction on:'' followed by a very large expression. In order to improve this usability, it could make sense to only do the auto-population if the user has just selected a variable, and otherwise leave it blank.

These barriers to starting show some room for improvement with RQ2 (Usability), but there are some reasonably simple interventions that could be made to improve usability.

\subsubsection{Evaluation Stepping}
\label{sec:log_ev_step}

The biggest change we made to the tool between the two deployments was to the way evaluation stepping worked, and so as expected there is a wide difference between the observed patterns in the two deployments. In the first deployment 33\% then 50\% then 0\% (\btref{c}) of students deleted evaluation steps in each induction respectively. This shows that students were frequently trying evaluation steps and then backtracking. Interestingly, when watching the student log replays, it was clear that most of the time the student had even chosen the correct step, but were still backtracking for lack of confidence. We can see that the second induction had the most backtracking, and this can perhaps be explained by the fact that it required the most evaluation steps to complete the proof. Students were simply losing confidence after a certain number of steps.

By complete contrast, in the second deployment, no student ever specifically deleted an evaluation step (\btref{d}). This is likely because in the second deployment, it required significantly more work in order to create an evaluation step. Students would need to identify the subexpression to step themselves, and then type out what the subexpression becomes. Once students have been through this process the students were clearly more confident in their written step.

\subsubsection{Recognizing Rewrites}
\label{sec:log_rewrites}

In each of the inductive steps, users also need to identify when to break from stepping and rewrite according to the inductive step or a lemma. The third and fourth inductions also introduce the requirement for the user to add a lemma in the inductive step. The third exercise explicitly tells the student that they will need the lemma, whereas the fourth exercise (which only existed on the second deployment) provides the lemma presented as an example proof, without stating that it is a lemma. 

Backtracking rates are significantly higher before the inductive hypothesis than after (\btref{e}). Across the three exercises, in the first deployment, the average student performed 32.7 backtracking actions before applying the inductive hypothesis, and only 7.1 backtracking actions after. This suggests that a lot of this backtracking could be down to students trying to find the correct place to put the inductive hypothesis.

One particularly interesting manifestation of this backtracking was students' opening up inner inductions. In the first deployment $44\%$ (\btref{f}) of students did this on the first induction exercise, but this number went down to $28\%$ on the second induction exercise (despite higher levels of overall backtracking), and none of them did it on the last one, suggesting that users quickly learned that opening an inner induction wasn't productive, and didn't go back to it. In the second deployment we did see an uptick in students attempting inner inductions for the final exercise (\btref{g}) - Peano multiplication, which we hypothesize is due to the fact that they weren't explicitly told to use the Peano addition lemma, so before students realized they can use that proof, they likely saw the inductive nature of addition inside the multiplication and decided to try an inner induction again.

\subsubsection{Log Methods and Exclusions} \label{sec:log_methods}
Along with the final proofs, we collected a list of actions that the student took as part of their submission. The original logs could not be directly analyzed because they were 1) too fine grained, and 2) lacked any information about the model at the time the action was taken. This information about the model would be necessary to work out whether an action was productive, and to work out when the student had reached each of our milestones. In order to summarize the logs, we replayed the logs in Hazel, recorded summarized actions, and recorded when the student had reached certain milestones.

For our first deployment there were 25 student submissions. Of these 25 submissions, 2 students did not attempt the proof questions. We were unable to replay a further 5 of the submissions due to bugs in the replay system, (which we fixed for the second deployment). This means in total we analyzed the logs of 18 submissions, these submissions all completed all exercises.

For our second deployment, there were 16 student submissions. One student submitted a pdf with screenshots of Hazel instead of the Hazel file itself; we were unable to replay one of the logs due to remaining Hazel bugs. One student completed the first induction but did not attempt the rest; one student was unable to finish the inductive step on the first question but completed all the other exercises. Thus, in total, we analyzed 14 submissions.

\subsection{Survey Results}
\label{sec:survey}

\begin{figure}
    \centering
    \includegraphics[width=\linewidth]{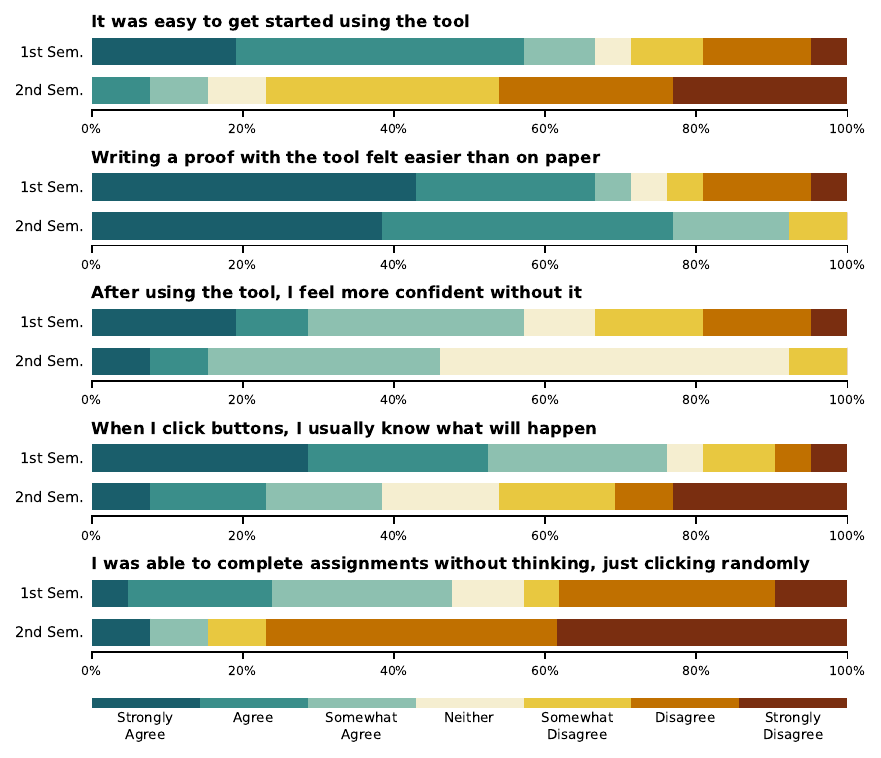}
    \caption{Survey results for the Hazel Prover.}
    \label{fig:sruvey_prover}
\end{figure}

In both of our two deployments, a mid-semester survey was given to the students asking for their level of agreement with statements relating to our design goals (\autoref{sec:ctns}). 21 students completed the survey in the undergraduate class and 13 students completed the survey in the graduate class. In both of these offerings students were given course credit to incentivize taking the survey.

\subsubsection{Evaluating Learning}
\label{sec:survey_learning}



In the first deployment, the high rate of students stating they could complete assignments by ``clicking randomly'' led us to conclude that the students may not have actually engaged enough to learn. This number decreased from 48\% in the first deployment to 15\% in the second deployment, suggesting our intervention to increase engagement (see \autoref{sec:evstep}) was successful.

The design iteration between the first prover deployment and the second did however lead to a decrease in students agreeing with ``When I click buttons, I usually know what will happen". We posit this decrease is expected due to the nature of the buttons in the second deployment, despite our design criterion \ctnref{ctn:predictable} preferring predictability. In the second deployment, one key button students used often was one checking the validity of an evaluation step when specified by-hand; this button must necessarily sometimes produce unexpected results, as the button telling students whether the rewrite is valid is the main way students receive feedback.

\subsubsection{Evaluating Transfer}
\label{sec:survey_transfer}

Improving knowledge and skill transfer was the core goal of our design iteration. The engagement with learning discussed in the previous section is clearly a prerequisite to transfer, as students that are not engaged will not learn enough to prove without the tool. Any attempt to improve transfer must therefore begin by improving engagement.

To see whether this improvement in engagement had an effect on students' perceptions of transfer, we also had an explicit survey question asking about students' confidence without the tool. In the second semester significantly fewer students stated they disagreed with this statement (though slightly fewer students agreed as well). We see this as an improvement in students' self-perception of transfer in the tool.

\subsubsection{Evaluating Usability}
\label{sec:survey_usability}

One aspect of usability that was supported strongly in both deployments was the ``easier than on paper" feedback, which a majority of students in both semesters supported. This suggests 1) that the stepper is indeed supporting students through a task that is otherwise difficult to do on paper (\ctnref{ctn:discover}), and 2) that the design of the prover did not introduce extra obligations or difficulties in its formalization (\ctnref{ctn:tedium}).

One stark change to survey results after our design iteration was the sharp jump in students disagreeing with statements about the ease-of-use of the prover. The related design criterion here is \ctnref{ctn:setup}. We attribute the majority of this issue to an oversight on our part: a lack of resources provided to students such as walkthrough videos or tutorials. The first iteration had the instructor demonstrate the tool in class, while in the second iteration students were tasked with learning to use the tool with only a minimal written description. With this lack of training, some second semester students struggled initially to use the prover and manipulate proof terms. For example, playing back student log data corroborates that some students struggled to discover that highlighting subexpressions was the way to step them (see \autoref{sec:log_begin_ind}). 

However, the ease of use feedback from students in the first semester support our usability goals. Coupled with the fact that our design iteration was relatively small (only needing to write out evaluation steps manually), we attribute much of the change to the difference of training provided to students, rather than a radically worse design. Providing greater instructional resources will be a key goal in later deployments, and our first-semester data supports the conclusion that students find the prover usable without too much support required.





\subsection{Exam Results}
\label{sec:exam_results}

\begin{figure}
    \centering
    \includegraphics[width=\linewidth]{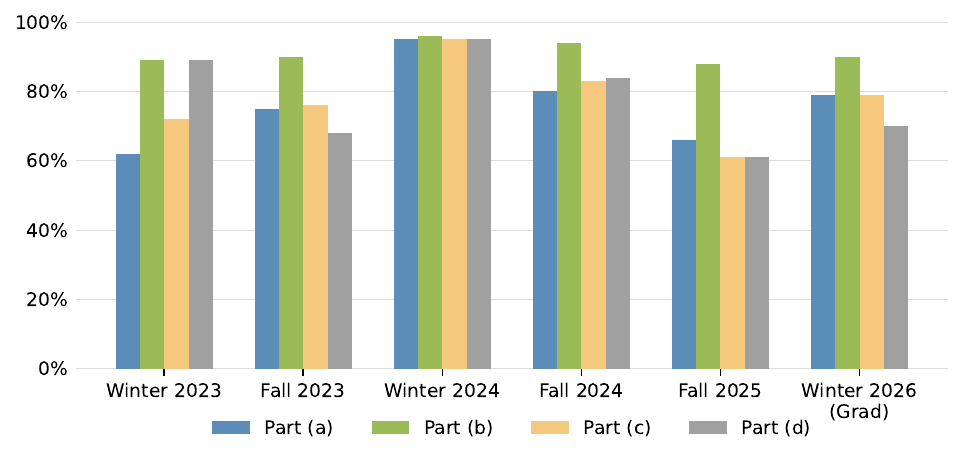}
    \caption{Average student score on each part of the induction question on the midterm for each semester. Hazel Prover was only used in Fall 2025. Students were asked to give (a) the inductive predicate, (b) the proof of the base case, (c) the inductive hypothesis, and (d) the proof of the inductive step.}
    \label{fig:scores}
\end{figure}

The average student exam results for each part of the induction question are shown in the chart in \autoref{fig:scores}. The four offerings of the undergraduate class prior to deployment are included for comparison. It should be noted that while all these exam questions followed the same four-part structure, the questions themselves were different and could vary in difficulty. The questions asked in the Fall 2025 semester and Winter 2026 semester (the two semesters we deployed the tool) were identical. The number of students taking the exam each semester is shown in \autoref{tab:enrollment}.

\begin{figure}
    \centering
    \includegraphics[width=\linewidth]{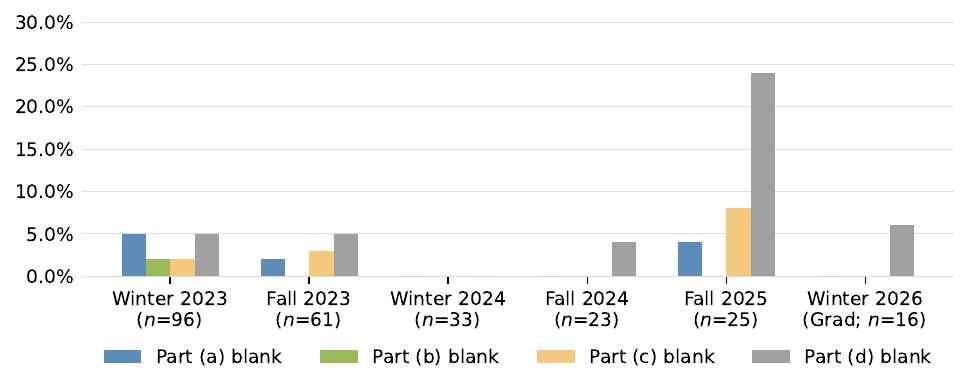}
    \caption{Proportion of students leaving each part of the induction question on the midterm exam blank. Hazel Prover was only used in Fall 2025. Each semester had a different enrollment with later semesters having lower enrollments.}
    \label{fig:blanks}
\end{figure}

The scores were visibly lower in the Fall 2025 semester, the first semester of using the tool, than in previous semesters. In the second deployment, however, the scores appear to have recovered closer to the normal range. It was also striking in \autoref{fig:blanks} that a significant proportion of students left the inductive step part of the question blank in the exam in the Fall 2025 semester, but that that number again recovered back to the normal range in Winter 2026.

Since the question varied semester-to-semester (except for Fall 2025 and Winter 2026 which had the same question), it is impossible to truly know whether students were performing worse in Fall 2025 (the semester we deployed the tool), or whether the question in Fall 2025 was just harder for students. It is, however, worth noting that except for part (a) in Winter 2023, students performed worse in Fall 2025 on each part than in any previous semester. This likely reflects the fact that the previous version of the homework assigned to students before the stepper had a much closer mapping to the four-part pen-and-paper question we asked them to answer in the exam.

While it is again difficult to compare Winter 2026 because it was a graduate course instead of an upper-level undergraduate course, the exam scores in that class appear to be similar to the earlier semesters of the undergraduate class. This could suggest that our intervention to make students write out steps (\autoref{sec:evstep}) improved students' performance in exams as it made this homework's engagement with the problem far more similar to the engagement on pen-and-paper. 

The fact that students are leaving the inductive step blank, suggests that they are missing the support (\ctnref{ctn:discover}) provided in the inductive step by the tool. Since in the Winter 2026 deployment, students were significantly less likely to leave the question blank, it suggests that the new pattern of engagement after the design iteration meant that after using the new version of the tool, students were more easily able to start the proof without support in the exam.

\begin{table}
    \centering
    \begin{tabular}{r|r}
        Semester & Number of students \\ \hline
        Winter 2023 & 96 \\
        Fall 2023 & 61 \\
        Winter 2024 & 33 \\
        Fall 2024 & 23 \\
        Fall 2025 & 25 \\
        Winter 2026 (graduate class) & 16 \\
    \end{tabular}
    \caption{Number of students taking the midterm each semester.}
    \label{tab:enrollment}
\end{table}

\subsection{Instructors' Perceptions of the Tool}

Instructors interact with the Prover by (1) demonstrating the prover in class in order to teach induction, (2) setting up the assignments for students, and responding to student questions about the tool, and (3) grading submissions from the tool. Each of these were done by a different instructor in the class. The tool author was not an instructor.

The instructor responsible for demonstrating the prover in the main lecture (in the first iteration only) for the class was the last author of this paper. This instructor reported a positive experience with the Prover. In prior semesters, induction proofs were done on a virtual whiteboard. This required spending time copying over unchanged code for each line of an equational proof, slowing down the lecture. By using the Prover, the instructor was able to cover more examples in the lecture. Explaining the user interface did not require significant effort, and there were only a few quick clarifying questions specific to the user interface in lecture.

The instructor responsible for setting up the assignment reported that ``setting up the definitions and theorem statements was not time consuming", but since it was a new tool they had to run through the exercises checking for bugs in the tool, which did take time. This suggests with some more polish and bug fixing we can achieve \ctnref{ctn:setup}. When asked whether they received many questions about how to use the tool, they replied, ``not many that I can remember", which indicates students also didn't have trouble setting up the tool, and perhaps validates that we have succeeded in \ctnref{ctn:curve}.

While Hazel Deriver does provide an autograder that can tell whether a proof is complete, a grader in the class checked through the proofs to check that no student had written a proof that was valid but convoluted, using meaningless steps or multiple inductions when only one was necessary. The grader reported that ``the structured nature of the stepper made it easier to follow the logical flow and verify whether each step was justified correctly, compared to deciphering handwritten arguments", and added that ``the stepper definitely reduced ambiguity, which helped with consistency in grading".

\subsection{Summary of Findings}
\label{sec:summary_of_findings}

\subsubsection{RQ1a (Learning)}

Through our log findings, we saw that students became more confident with the tool as they were completing exercises, evidenced in particular by the reduction in backtracking actions over time. While some of these backtracking reductions are perhaps evidence of students becoming more familiar with the tool itself (such as backtracking at the start), some of the backtracking reductions point to genuine lessons in learning how to prove structural inductions (such as learning to use the IH instead of trying another inner induction). We saw that students did feel supported by the tool in their learning, as evidenced by the fact that a large majority of students reported that it was easier to use than pen-and-paper.

We saw a big difference in levels of engagement between the two deployments, evidenced by the students' response to the survey question asking about ``clicking randomly''. This is a key prerequisite to learning, and suggests that students' actual learning of induction may have been hampered in the first deployment, but that we fixed this issue in the second deployment.

\subsubsection{RQ1b (Transfer)}

On top of students' engagement with the tool improving, the fact that the exam scores improved also suggest that the increased engagement in the second deployment increased transfer. The clearest signal was the proportion of students who left the exam question blank. This was significantly higher in the first deployment than the second, suggesting the second tool at least led to much better confidence on pen-and-paper. Care must be taken in the comparison, however, since the two classes had different cohorts (one was upper-level undergraduate level, and the other was graduate level). However, scores on other common problems on the two exams suggested that the graduate students were not otherwise substantially more capable than the undergraduate students with this material.

These results are also reflected in students' self-perception of transfer, as in the second deployment, significantly fewer students disagreed with the statement that they felt more confident without the tool after using the tool.

The fact that we have such a big difference between the two deployments means we can say not only that our tool supports transfer, but we can pinpoint exactly what aspects of the tool support transfer, specifically the parts of the tool that force engagement in a similar manner to how the paper proof would force engagement.

\subsubsection{RQ2 (Usability)}

The survey reported that a large majority of students in both deployments found using Hazel Prover to be easier than a pen-and-paper proof, however they had some issues getting started, particularly on the second iteration where students were not given any instruction. In the survey for the first deployment a large majority of students agreed with the statement that it was easy to get started with the tool, whereas in the second deployment a large majority disagreed. Student comments in the second deployment pointed to the fact that it was not clear how to begin by selecting; and this was borne out in log observations, where students would frequently try using other parts of the interface before selecting and starting an induction, though these usability issues decreased quickly, which was observed in the decrease in students' backtracing before starting an induction.

\subsection{Threats to Validity}

While deploying the tool in real classes for real assignments provides a level of authenticity to our evaluation that wouldn't be available in tightly-controlled user studies, it also introduces several factors that we cannot control that hamper fair comparisons. The two deployments were completed in different classes at different levels, with one undergraduate and one graduate class. The two classes, with different syllabuses were also inconsistent in the amount of training on our tool that was given to students. Our assignments were specific to these two classes at this institution, and may not generalize to other classes. 

The bugs in log replay create several threats to validity, since in the first deployment we were only able to replay 18 of the 25 student logs, our sample size is even smaller, and we run the risk that the other students' logs may contain a pattern we are unable to observe. There is also a risk that some meaningful pattern (e.g. extreme use of undo/redo) could be the cause of us being unable to replay the student logs, giving us a false sense of their interaction.

%% file: sections/50_related_work.tex
\section{Related Work}
\label{sec:related_work}

Hazel Prover is a classroom proof assistant for teaching structural induction. We will begin by discussing the use of proof assistants in the classroom in general, and then narrow in to discussing classroom proof assistants specific to structural induction. Some authors have taken full-scale proof assistants directly into the classroom, and others have adapted these proof assistants for education, either by replacing the tactics with more education-friendly tactics, or by creating a new interface designed for the classroom being used in the classroom. Two recent surveys of deployments of classroom proof assistants have been completed by \citet{keenan_learner-centered_nodate} and \citet{minh_proof_2025}.

Popular examples of full-scale proof assistants used in classrooms include Rocq \cite{coq_development_team_coq_2024}, Lean \cite{leanprover} and Isabelle \cite{nipkow_isabellehol_2002}, among others. These proof assistants are used in research and industry in the formalization of mathematics and engineering projects \cite{chen2025review}. There have been many attempts at introducing these tools into the classroom, dating all the way back to the introduction of MIZAR to the classroom in 1975 \cite{matuszewski2005mizar}. We have listed some more notable examples in \autoref{tab:full}.

\begin{table}[t]
\begin{tabular}{|l|l|}
    \hline \textbf{Theorem Prover} & \textbf{Examples} \\
    \hline
    Rocq & 
    \citet{10.1145/3758317.3759679}, \citet{kerjean_utilisation_2022},  \\
    \cite{coq_development_team_coq_2024} & 
    \citet{knobelsdorf_theorem_2017}, \citet{gallego_arias_jscoq_2017}, \\
    & \citet{henz_teaching_2011}, \citet{hendriks_teaching_2010}, \\
    & \citet{pierce_lambda_2009}, \citet{delahaye_coq_2005}\\ 
    \hline
    Lean & \citet{10.1145/3758317.3759679}, \citet{kerjean_utilisation_2022}, \\
    \cite{leanprover} & \citet{thoma_learning_2022}, \citet{avigad_learning_2019} \\
    \hline
    Isabelle  & \citet{jacobsen_exams_2023}, \citet{nipkow_teaching_2012}\\ 
    \cite{nipkow_isabellehol_2002} & \\
    \hline
\end{tabular}
\caption{Experiences using full-scale proof assistants with students.}
\label{tab:full}
\vspace{-10pt}
\end{table}

While using a tool from industry and research may make the tool more ``authentic'' for students, these proof assistants' power and generality can make the experience worse for teaching. In particular, deployments of these proof assistants have found students struggle with writing the tools' complex syntax, and understanding confusing error messages \cite{10.1145/3758317.3759679}. On top of these difficulties in using the tool, these proof assistants may also automate too much for the user, preventing them from practicing important parts of the proof \cite{bartzia2023proof}. Deployments have also shown that even if students can learn how to use one of these tools, this does not always transfer to them being able to prove without the tool \cite{knobelsdorf_theorem_2017}, especially since the syntax and layout of an automated proof can be very different to how it would be structured in a textbook \cite{bohne_learning_2018}.

In adapting proof assistants for education, some authors have taken these full-scale proof assistants and replaced their tactics and interface with tactics and interfaces more conducive to education. The most mature such system is {\it Waterproof}~\cite{wemmenhove_waterproof_2024} which is built on top of Rocq, and has been used in teaching real analysis for six years \cite{wemmenhove2025waterproof}. The developers of Waterproof took steps to create a closer correspondence between Waterproof proofs and written proofs, including adding extra signposting tactics, which full-scale proof assistants would usually not require the user to write, but that make the final typed proof's structure more closely resemble a written one, in order to better facilitate transfer. A few projects have recently adapted this approach to Lean as well \cite{massot2024teaching,minh2025lean}. While Waterproof is a very mature system now and has certainly shown this approach to be viable, it is our opinion that tactic-based systems still place a heavy syntax burden on the user, which we are keen to avoid. We would also like students to be able to directly manipulate equations instead of just working on a console. Hazel Prover is notable in that the only symbolic notation needed is exactly the notation of the Hazel programming language; all other interactions are via graphical user interface elements that require more recognition and less recall.


There has also been a long tradition of authors creating bespoke classroom tools for teaching natural deduction (as opposed to our goal of teaching equational and inductive reasoning), from which we can draw inspiration. One of the earliest such works is the CMU Proof Tutor~\cite{scheines1994computer}, which laid out proofs in a tree of goals, and allowed students to click to apply rules to applicable goals. More recent systems include Holbert~\cite{oconnor_holbert_2022} which takes a literate approach, allowing proofs to be embedded between paragraphs, and letting users choose between proof tree and structured text proof formats. Edukera~\cite{rognier_presentation_2016} is another interface that allows proving natural deduction and other high school math topics all using point-and-click. Edukera has been deployed in several French universities \cite{kerjean_utilisation_2022}. Finally, Hazel Deriver is another classroom proof assistant, this one for building derivation trees that, like Hazel Prover, is integrated into Hazel~\cite{zhong_hazel_2025}. Classroom use was not previously reported for Hazel Deriver. We discuss it further in \autoref{sec:discussion}.


The most similar projects to the Hazel Prover that focus on the same kinds of proof problems are LogInd~\cite{lodder_providing_2020} and Emmy~\cite{xu_emmy_nodate}. LogInd has been deployed in an online course at the Open University, and it has been used regularly since 2022 \cite{gerdes_2025_17181795}. We are not aware of any classroom deployments of Emmy. Both tools, like ours, provide scaffolded user interfaces for structural induction, where the user can add their own cases, and both tools provide equational reasoning where the user can prove an equality by writing out a chain of equality steps and having the steps automatically checked. LogInd has a strong deployment history, and has implemented a fully-featured hint system for when students get stuck. Because LogInd targets a specific guided exercise format, the interface is more rigid than ours, with students only able to prove a single structural induction, and no facility for the student to choose when to start an induction, to decide to nest inductions, or to decide what data types to induct over. The proof is also fully set up by the instructor, with students unable to provide lemmas or edit the specification. Emmy offers the user more flexibility to decide how to structure the proof, but has a less ergonomic interface, and as far as we are aware, does not have a history of deployment. In addition, unlike Hazel Prover, neither of these tools integrate the induction into a programming environment, so in order to prove facts about functions, one must first provide an axiomatic semantics of the functions used. Neither tool has reported successful transfer to on-paper proofs, and our analysis goes more in-depth into student usage patterns.

%% file: sections/60_conclusion.tex
\section{Discussion and Future Work}
\label{sec:discussion}




\subsection{Achieving Transfer}


In this work we identified a set of design criteria for transfer to pen-and-paper, and showed that ensuring closeness of mapping between engagement with the tool and with pen-and-paper proof, improves students' confidence on paper after using the tool. This gives us a way to design a custom proof assistant that directly facilitates transfer, avoiding prior work's need to devote extra time to specifically teaching students to transfer their knowledge~\cite{bohne_learning_2018}.

In our first class deployment, while the students were able to use the tool in order to complete the assignment, we observed difficulties with transferring these skills to pen-and-paper. The key observations were that exam results in the first deployment had a very high rate of students leaving the question blank, and that many students' survey responses explicitly disagreed with the proposition:
``After using the tool, I feel more confident without it.''

Given these observations, we then identified a lack of engagement as the root cause for the lack of transfer, for two reasons. Firstly, an alarmingly high number of students said they were able to complete exercises randomly (\autoref{sec:survey_learning}), suggesting that our design was not meeting \ctnref{ctn:engagement}. Secondly, we saw that students had responded much more positively to the survey questions about transfer for the other tool we were using in class, Hazel Deriver~\cite{zhong_hazel_2025}, which is designed to emphasize engagement more. 

One of the Deriver's key advantages is likely how closely the task of writing a derivation tree in the deriver and the task of writing the same tree on pen-and-paper map on to each other. The design explicitly avoids any automation \cite{zhong_hazel_2025}, forcing users to write out every layer of the tree by hand, even though much of this could be automated, and even asks users to add the correct number of premises for each rule they invoke. This means users practice every skill they would need to be able to use when drawing out a tree, while still getting instantaneous feedback on their tree.

\citet{knobelsdorf_theorem_2017} theorized that the Rocq proof skills their students learned didn't easily transfer to pen-and-paper because of how different a Rocq proof and a proof on pen-and-paper are presented. While we focused, in our design, on a closeness of mapping of notation used in Hazel Prover and on paper, we have also seen through experience that we need to strengthen this criterion to a closeness of mapping of engagements (\ctnref{ctn:engagement2}) used in the tool and on paper.

We therefore iterated the design to force students to engage more with writing out evaluation steps (\autoref{sec:evstep}). This more closely matched that fact that when writing proofs on paper, students would need to perform the evaluation themselves in their head. This design successfully forced engagement, and using log data we observed students taking more ownership of the steps they had written (\autoref{sec:log_ev_step}). We also saw a large reduction in the proportion of students who said they were able to complete exercises randomly in Hazel Prover.

This improvement in engagement also translated to an improvement in transfer. We saw a significant decrease in the proportion of students who explicitly disagreed with the idea that the tool made them more confident on paper, and we saw a significant decrease in the proportion of students who left the exam question blank in the second deployment (\autoref{sec:exam_results}), despite scores for students who did not leave it blank remaining roughly similar, suggesting we do have comparable cohorts for the exams.

Thus we have shown that by closely mapping a design to pen-and-paper proofs, both in the notation (\autoref{ctn:notation}) and engagement (\autoref{ctn:engagement2}), students can learn to write proofs on a proof assistant, with all the feedback and scaffolding it provides, while still having that skill transfer to confidence on pen-and-paper.




\subsection{Engagement Without Aiming for Transfer}

While we made transfer to pen-and-paper proof an explicit learning goal of our proof assistant, and have taken steps to modify our design to force transfer to pen-and-paper, it is worth also discussing the implications of these findings for contexts where transfer to pen-and-paper is not a requirement.

The importance of transfer to pen-and-paper may simply reflect the way we currently assess students, and it is worth considering whether it may not reflect the kinds of exercises we would ideally like to give students. In particular, all the examples we gave students in \autoref{sec:class} were small toy examples on lists, since the evaluations required to do those proofs on lists made the proof about as long as we could reasonably ask students to produce on paper, or using this version of the Hazel Prover. If we instead allowed for more automation, this support might allow students to reach questions that require more planning, or feel closer to real-world applications.

Indeed, as LLM and proof assistant technologies allow automation of the simpler tasks in proofs, it may be the case that teaching students to be able to plan proofs, and to write their own specifications, is a far more valuable tool than being able to produce simple proofs like these.

However, not measuring transfer comes at a significant disadvantage that it is harder to measure the learning outcome. Without explicitly measuring students' transfer to pen and paper, we may not have taken notice of the problems with engagement in the first deployment. Being able to produce a pen-and-paper proof is a hard skill we can grade. Learning the logic behind proofs is a soft skill, and it is hard to assess the extent to which a student has learned it, or whether all the tool's automation could be hiding some fundamental misunderstandings in the student.

\subsection{Design Appraisal and Critique}

Our design for a point-and-click prover built on single-step evaluation and extended with subterm selection and rewriting has shown to be an effective design for a classroom proof assistant that can prove properties of real functional code using structural induction. In particular, the stepper, after some changes to increase engagement, lends itself naturally to a similar layout and engagement to that seen on pen-and-paper proofs, and does not require much training.

In \autoref{sec:log_begin_ind} we noted some usability issues with students struggling to figure out how to start using the tool. When students first open the tool, particularly if they had not had the tool demonstrated to them, there was no clear affordance for starting the proof, and it was unintuitive that you could select a variable to begin an induction on that variable. In order to make it clearer for students in the future we should find ways to help students past that initial interaction hurdle, perhaps by leaving a ``proof hole'' with instructions.

We demonstrated the tool's use on structural induction proofs, but in future work we would like to consider extending it to other domains. In particular we should consider how to extend it to support other logical constructs such as implication, which could allow us to go further in programming languages courses and prove properties such as preservation.

Another interesting direction for future work would be to extend the stepper design to support other mathematical topics where equational reasoning is important. We could do a similar investigation to see whether our mapping-of-engagement principle also applies to those contexts.

\section*{Data Availability Statement}

The supplemental material includes additional details from the study and a pre-built implementation of Hazel Prover that the reviewers can use to follow along with the examples in the paper. We plan to submit an artifact with this information as well as anonymized study data and analysis scripts. Hazel Prover will be released as an open source project and we plan to support instructors who are interested in deploying it into their own classes.

%% file: sections/99_references.bib
@article{minh_proof_2025,
	title = {Proof {Assistants} for {Teaching}: a {Survey}},
	volume = {419},
	issn = {2075-2180},
	shorttitle = {Proof {Assistants} for {Teaching}},
	url = {http://arxiv.org/abs/2505.13472},
	doi = {10.4204/EPTCS.419.1},
	urldate = {2025-06-09},
	journal = {Electronic Proceedings in Theoretical Computer Science},
	author = {Minh, Frédéric Tran and Gonnord, Laure and Narboux, Julien},
	month = may,
	year = {2025},
	note = {arXiv:2505.13472 [cs]},
	pages = {1--27},
}

@inproceedings{keenan_learner-centered_nodate,
	title = {Learner-{Centered} {Design} {Criteria} for {Classroom} {Proof} {Assistants}},
    booktitle = {Human Aspects of Types and Reasoning Assistants (HATRA)},
    year = {2024},
	language = {en},
    url = {https://hazel.org/papers/learner_centered_proofs_hatra_2024.pdf},
	author = {Keenan, Matthew and Omar, Cyrus},
}

@inproceedings{knobelsdorf_theorem_2017,
	address = {Tacoma Washington USA},
	title = {Theorem {Provers} as a {Learning} {Tool} in {Theory} of {Computation}},
	isbn = {978-1-4503-4968-0},
	url = {https://dl.acm.org/doi/10.1145/3105726.3106184},
	doi = {10.1145/3105726.3106184},
	language = {en},
	urldate = {2025-12-09},
	booktitle = {Proceedings of the 2017 {ACM} {Conference} on {International} {Computing} {Education} {Research}},
	publisher = {ACM},
	author = {Knobelsdorf, Maria and Frede, Christiane and Böhne, Sebastian and Kreitz, Christoph},
	month = aug,
	year = {2017},
	pages = {83--92},
}

@article{lodder_providing_2020,
	title = {Providing {Hints}, {Next} {Steps} and {Feedback} in a {Tutoring} {System} for {Structural} {Induction}},
	volume = {313},
	issn = {2075-2180},
	url = {http://arxiv.org/abs/2002.12552v1},
	doi = {10.4204/EPTCS.313.2},
	language = {en},
	urldate = {2025-12-09},
	journal = {Electronic Proceedings in Theoretical Computer Science},
	author = {Lodder, Josje and Heeren, Bastiaan and Jeuring, Johan},
	month = feb,
	year = {2020},
	pages = {17--34},
}

@misc{oconnor_holbert_2022,
	title = {Holbert: {Reading}, {Writing}, {Proving} and {Learning} in the {Browser}},
	shorttitle = {Holbert},
	url = {http://arxiv.org/abs/2210.11411},
	urldate = {2023-10-16},
	publisher = {arXiv},
	author = {O'Connor, Liam and Amjad, Rayhana},
	month = oct,
	year = {2022},
}

@article{bohne_learning_2018,
	title = {Learning how to {Prove}: {From} the {Coq} {Proof} {Assistant} to {Textbook} {Style}},
	volume = {267},
	issn = {2075-2180},
	shorttitle = {Learning how to {Prove}},
	url = {http://arxiv.org/abs/1803.01466},
	doi = {10.4204/EPTCS.267.1},
	urldate = {2024-02-17},
	journal = {Electronic Proceedings in Theoretical Computer Science},
	author = {Böhne, Sebastian and Kreitz, Christoph},
	month = mar,
	year = {2018},
	pages = {1--18},
}

@inproceedings{rognier_presentation_2016,
	address = {Saint-Malo, France},
	title = {Présentation de la plateforme edukera},
	url = {https://hal.science/hal-01333606},
	urldate = {2024-02-17},
	booktitle = {Vingt-septièmes {Journées} {Francophones} des {Langages} {Applicatifs} ({JFLA} 2016)},
	author = {Rognier, Benoit and Duhamel, Guillaume},
	editor = {Signoles, Julien},
	month = jan,
	year = {2016},
}

@article{xu_emmy_nodate,
	title = {Emmy: {A} {Proof} {Assistant} for {Reasoning} about {Programs}},
	language = {en},
    year = {2019},
	author = {Xu, Junfeng and Drossopoulou, Sophia}
}

@article{kerjean_utilisation_2022,
	title = {Utilisation des assistants de preuves pour l'enseignement en {L1}},
	volume = {174},
	url = {https://hal.science/hal-03979238},
	urldate = {2024-06-05},
	journal = {Gazette des Mathématiciens},
	author = {Kerjean, Marie and Le Roux, Frédéric and Massot, Patrick and Mayero, Micaela and Mesnil, Zoé and Modeste, Simon and Narboux, Julien and Rousselin, Pierre},
	month = oct,
	year = {2022},
}

@article{wemmenhove_waterproof_2024,
	title = {Waterproof: {Educational} {Software} for {Learning} {How} to {Write} {Mathematical} {Proofs}},
	volume = {400},
	issn = {2075-2180},
	shorttitle = {Waterproof},
	url = {http://arxiv.org/abs/2211.13513v2},
	doi = {10.4204/EPTCS.400.7},
	language = {en},
	urldate = {2024-06-05},
	journal = {Electronic Proceedings in Theoretical Computer Science},
	author = {Wemmenhove, Jelle and Arends, Dick and Beurskens, Thijs and Bhaid, Maitreyee and McCarren, Sean and Moraal, Jan and Rivera Garrido, Diego and Tuin, David and Vassallo, Malcolm and Wils, Pieter and Portegies, Jim},
	month = apr,
	year = {2024},
	pages = {96--119},
}

@article{thoma_learning_2022,
	title = {Learning about {Proof} with the {Theorem} {Prover} {LEAN}: the {Abundant} {Numbers} {Task}},
	volume = {8},
	issn = {2198-9753},
	shorttitle = {Learning about {Proof} with the {Theorem} {Prover} {LEAN}},
	url = {https://doi.org/10.1007/s40753-021-00140-1},
	doi = {10.1007/s40753-021-00140-1},
	language = {en},
	number = {1},
	urldate = {2024-06-11},
	journal = {International Journal of Research in Undergraduate Mathematics Education},
	author = {Thoma, Athina and Iannone, Paola},
	month = apr,
	year = {2022},
	pages = {64--93},
}

@incollection{avigad_learning_2019,
	address = {Cham},
	title = {Learning {Logic} and {Proof} with an {Interactive} {Theorem} {Prover}},
	isbn = {978-3-030-28483-1},
	url = {https://doi.org/10.1007/978-3-030-28483-1_13},
	language = {en},
	urldate = {2024-06-11},
	booktitle = {Proof {Technology} in {Mathematics} {Research} and {Teaching}},
	publisher = {Springer International Publishing},
	author = {Avigad, Jeremy},
	editor = {Hanna, Gila and Reid, David A. and de Villiers, Michael},
	year = {2019},
	doi = {10.1007/978-3-030-28483-1_13},
	pages = {277--290},
}

@article{gallego_arias_jscoq_2017,
	title = {{jsCoq}: {Towards} {Hybrid} {Theorem} {Proving} {Interfaces}},
	volume = {239},
	issn = {2075-2180},
	shorttitle = {{jsCoq}},
	url = {http://arxiv.org/abs/1701.07125},
	doi = {10.4204/EPTCS.239.2},
	language = {en},
	urldate = {2024-06-12},
	journal = {Electronic Proceedings in Theoretical Computer Science},
	author = {Gallego Arias, Emilio Jesús and Pin, Benoît and Jouvelot, Pierre},
	month = jan,
	year = {2017},
	pages = {15--27},
}

@article{byrne_student_2018,
	title = {Student {Interpretations} of {Written} {Comments} on {Graded} {Proofs}},
	volume = {4},
	issn = {2198-9753},
	url = {https://doi.org/10.1007/s40753-017-0059-0},
	doi = {10.1007/s40753-017-0059-0},
	language = {en},
	number = {2},
	urldate = {2026-01-01},
	journal = {International Journal of Research in Undergraduate Mathematics Education},
	author = {Byrne, Martha and Hanusch, Sarah and Moore, Robert C. and Fukawa-Connelly, Tim},
	month = jul,
	year = {2018},
	pages = {228--253},
}

@article{moore_mathematics_2016,
	title = {Mathematics {Professors}’ {Evaluation} of {Students}’ {Proofs}: {A} {Complex} {Teaching} {Practice}},
	volume = {2},
	issn = {2198-9753},
	shorttitle = {Mathematics {Professors}’ {Evaluation} of {Students}’ {Proofs}},
	url = {https://doi.org/10.1007/s40753-016-0029-y},
	doi = {10.1007/s40753-016-0029-y},
	language = {en},
	number = {2},
	urldate = {2026-01-02},
	journal = {International Journal of Research in Undergraduate Mathematics Education},
	author = {Moore, Robert C.},
	month = jul,
	year = {2016},
	pages = {246--278},
}

@article{miller_how_2018,
	title = {How mathematicians assign points to student proofs},
	volume = {49},
	issn = {0732-3123},
	url = {https://www.sciencedirect.com/science/article/pii/S0732312316301687},
	doi = {10.1016/j.jmathb.2017.03.002},
	urldate = {2026-01-02},
	journal = {The Journal of Mathematical Behavior},
	author = {Miller, David and Infante, Nicole and Weber, Keith},
	month = mar,
	year = {2018},
	pages = {24--34},
}

@article{delahaye_coq_2005,
	title = {Coq, un outil pour l'enseignement. {Une} expérience avec les étudiants du {DESS} {Développement} de logiciels srs},
	volume = {24},
	url = {https://doi.org/10.3166/tsi.24.1139-1160},
	doi = {10.3166/TSI.24.1139-1160},
	number = {9},
	journal = {Tech. Sci. Informatiques},
	author = {Delahaye, David and Jaume, Mathieu and Prevosto, Virgile},
	year = {2005},
	pages = {1139--1160},
}

@article{hendriks_teaching_2010,
	title = {Teaching logic using a state-of-the-art proof assistant},
	journal = {Acta Didactica Napocensia},
	author = {Hendriks, Maxim and Kaliszyk, Cezary and Raamsdonk, F van and Wiedijk, Freek},
	year = {2010},
}

@inproceedings{henz_teaching_2011,
	address = {Berlin, Heidelberg},
	title = {Teaching {Experience}: {Logic} and {Formal} {Methods} with {Coq}},
	isbn = {978-3-642-25379-9},
	shorttitle = {Teaching {Experience}},
	doi = {10.1007/978-3-642-25379-9_16},
	language = {en},
	booktitle = {Certified {Programs} and {Proofs}},
	publisher = {Springer},
	author = {Henz, Martin and Hobor, Aquinas},
	editor = {Jouannaud, Jean-Pierre and Shao, Zhong},
	year = {2011},
	pages = {199--215},
}

@article{jacobsen_exams_2023,
	title = {On {Exams} with the {Isabelle} {Proof} {Assistant}},
	volume = {375},
	issn = {2075-2180},
	url = {http://arxiv.org/abs/2303.05866v1},
	doi = {10.4204/EPTCS.375.6},
	language = {en},
	urldate = {2024-06-20},
	journal = {Electronic Proceedings in Theoretical Computer Science},
	author = {Jacobsen, Frederik Krogsdal and Villadsen, Jørgen},
	month = mar,
	year = {2023},
	pages = {63--76},
}

@article{guzdial_software-realized_1994,
	title = {Software-{Realized} {Scaffolding} to {Facilitate} {Programming} for {Science} {Learning}},
	volume = {4},
	url = {https://doi.org/10.1080/1049482940040101},
	doi = {10.1080/1049482940040101},
	number = {1},
	journal = {Interact. Learn. Environ.},
	author = {Guzdial, Mark},
	year = {1994},
	pages = {1--44},
}

@misc{coq_development_team_coq_2024,
	title = {The {Coq} {Proof} {Assistant}},
	url = {https://doi.org/10.5281/zenodo.11551307},
	publisher = {Zenodo},
	author = {Coq Development Team, The},
	month = jun,
	year = {2024},
	doi = {10.5281/zenodo.11551307},
	note = {Version Number: 8.19},
}

@inproceedings{leanprover,
  title = {The Lean 4 Theorem Prover and Programming Language},
  author = {de Moura, Leonardo and Ullrich, Sebastian},
  year = {2021},
  isbn = {978-3-030-79875-8},
  publisher = {Springer-Verlag},
  address = {Berlin, Heidelberg},
  url = {https://doi.org/10.1007/978-3-030-79876-5_37},
  doi = {10.1007/978-3-030-79876-5_37},
  booktitle = {Automated Deduction – CADE 28: 28th International Conference on Automated Deduction, Virtual Event, July 12–15, 2021, Proceedings},
  pages = {625–635},
  numpages = {11}
}

@book{nipkow_isabellehol_2002,
	title = {Isabelle/{HOL}: a proof assistant for higher-order logic},
	publisher = {Springer},
	author = {Nipkow, Tobias and Wenzel, Markus and Paulson, Lawrence C},
	year = {2002},
}

@inproceedings{nipkow_teaching_2012,
	address = {Berlin, Heidelberg},
	title = {Teaching {Semantics} with a {Proof} {Assistant}: {No} {More} {LSD} {Trip} {Proofs}},
	isbn = {978-3-642-27940-9},
	shorttitle = {Teaching {Semantics} with a {Proof} {Assistant}},
	doi = {10.1007/978-3-642-27940-9_3},
	language = {en},
	booktitle = {Verification, {Model} {Checking}, and {Abstract} {Interpretation}},
	publisher = {Springer},
	author = {Nipkow, Tobias},
	editor = {Kuncak, Viktor and Rybalchenko, Andrey},
	year = {2012},
	pages = {24--38},
}

@article{stylianides_research_2017,
	title = {Research on the teaching and learning of proof: {Taking} stock and moving forward},
	journal = {Compendium for research in mathematics education},
	author = {Stylianides, Gabriel J and Stylianides, Andreas J and Weber, Keith},
	year = {2017},
	pages = {237--266},
}

@article{omar_live_2019,
	title = {Live functional programming with typed holes},
	volume = {3},
	url = {https://doi.org/10.1145/3290327},
	doi = {10.1145/3290327},
	number = {POPL},
	journal = {Proc. ACM Program. Lang.},
	author = {Omar, Cyrus and Voysey, Ian and Chugh, Ravi and Hammer, Matthew A.},
	year = {2019},
	pages = {14:1--14:32},
}

@article{pierce_lambda_2009,
	title = {Lambda, the ultimate {TA}},
	volume = {44},
	number = {9},
	journal = {ACM Sigplan Notices},
	author = {Pierce, Benjamin C},
	year = {2009},
	pages = {121--122},
}

@misc{zhong_hazel_2025,
	title = {Hazel {Deriver}: {A} {Live} {Editor} for {Constructing} {Rule}-{Based} {Derivations}},
	shorttitle = {Hazel {Deriver}},
	url = {http://arxiv.org/abs/2506.10781},
	doi = {10.48550/arXiv.2506.10781},
	urldate = {2026-01-05},
	publisher = {arXiv},
	author = {Zhong, Zhiyao and Omar, Cyrus},
	month = jun,
	year = {2025},
	note = {arXiv:2506.10781 [cs]},
}

@article{wadler2015propositions,
  title={Propositions as types},
  author={Wadler, Philip},
  journal={Communications of the ACM},
  volume={58},
  number={12},
  pages={75--84},
  year={2015},
  publisher={ACM New York, NY, USA}
}

@misc{noauthor_algebrite_nodate,
	title = {Algebrite},
    year = {2016},
    author={Davide Della Casa},
	url = {http://algebrite.org/},
	urldate = {2026-01-09},
}

@inproceedings{10.1145/3758317.3759679,
author = {Mahinpei, Romina and Horta Ribeiro, Manoel and Milano, Mae},
title = {Interactive Theorem Provers for Proof Education},
year = {2025},
isbn = {9798400721427},
publisher = {Association for Computing Machinery},
address = {New York, NY, USA},
url = {https://doi.org/10.1145/3758317.3759679},
doi = {10.1145/3758317.3759679},
booktitle = {Proceedings of the 2025 ACM SIGPLAN International Symposium on SPLASH-E},
pages = {24–41},
numpages = {18},
location = {Singapore, Singapore},
series = {SPLASH-E '25}
}

@article{chen2025review,
  title={A Review on Mechanical Proving and Formalization of Mathematical Theorems},
  author={Chen, Si and Yu, Wensheng and Dou, Guowei and Zhang, Qimeng},
  journal={IEEE Access},
  year={2025},
  publisher={IEEE}
}

@article{bartzia2023proof,
  title={Proof assistants for undergraduate mathematics education: Elements of an a priori analysis},
  author={Bartzia, Evmorfia-Iro and Beffara, Emmanuel and Meyer, Antoine and Narboux, Julien},
  year={2023}
}

@article{wemmenhove2025waterproof,
  title={Waterproof: Transforming a proof assistant into an educational tool},
  author={Wemmenhove, Aalt Jelle},
  year={2025}
}

@inproceedings{massot2024teaching,
  title={Teaching mathematics using lean and controlled natural language},
  author={Massot, Patrick},
  booktitle={15th International Conference on Interactive Theorem Proving (ITP 2024)},
  pages={27--1},
  year={2024},
  organization={Schloss Dagstuhl--Leibniz-Zentrum f{\"u}r Informatik}
}

@inproceedings{minh2025lean,
  title={A Lean-based Language for Teaching Proof in High School},
  author={Minh, Fr{\'e}d{\'e}ric Tran and Gonnord, Laure and Narboux, Julien},
  booktitle={International Conference on Intelligent Computer Mathematics},
  pages={447--467},
  year={2025},
  organization={Springer}
}

@article{matuszewski2005mizar,
  title={Mizar: the first 30 years},
  author={Matuszewski, Roman and Rudnicki, Piotr},
  journal={Mechanized mathematics and its applications},
  volume={4},
  number={1},
  pages={3--24},
  year={2005}
}

@article{scheines1994computer,
  title={Computer environments for proof construction},
  author={Scheines, Richard and Sieg, Wilfried},
  journal={Interactive Learning Environments},
  volume={4},
  number={2},
  pages={159--169},
  year={1994},
  publisher={Taylor \& Francis}
}

@article{e0c7fdf9-b8a2-3d03-96c4-fbf15b9d2be0,
 ISSN = {02280671},
 URL = {https://www.jstor.org/stable/27091135},
 author = {Alon Pinto and Ronnie Karsenty},
 journal = {For the Learning of Mathematics},
 number = {1},
 pages = {pp. 22--27},
 publisher = {FLM Publishing Association},
 title = {NORMS OF PROOF IN DIFFERENT PEDAGOGICAL CONTEXTS},
 urldate = {2026-03-02},
 volume = {40},
 year = {2020}
}

@article{dawkins2017values,
  title={Values and norms of proof for mathematicians and students},
  author={Dawkins, Paul Christian and Weber, Keith},
  journal={Educational Studies in Mathematics},
  volume={95},
  number={2},
  pages={123--142},
  year={2017},
  publisher={Springer}
}

@inproceedings{clements2001modeling,
  title={Modeling an algebraic stepper},
  author={Clements, John and Flatt, Matthew and Felleisen, Matthias},
  booktitle={European symposium on programming},
  pages={320--334},
  year={2001},
  organization={Springer}
}

@misc{gerdes_2025_17181795,
  author       = {Gerdes, Alex and
                  Keuning, Hieke and
                  Swierstra, Wouter},
  title        = {Taalblaat - Festschrift for Johan Jeuring},
  month        = sep,
  year         = 2025,
  publisher    = {Zenodo},
  doi          = {10.5281/zenodo.17181795},
  url          = {https://doi.org/10.5281/zenodo.17181795},
}

@incollection{paulinmohring:hal-01094195,
  TITLE = {{Introduction to the Calculus of Inductive Constructions}},
  AUTHOR = {Paulin-Mohring, Christine},
  URL = {https://inria.hal.science/hal-01094195},
  BOOKTITLE = {{All about Proofs, Proofs for All}},
  EDITOR = {Bruno Woltzenlogel Paleo and David Delahaye},
  PUBLISHER = {{College Publications}},
  SERIES = {Studies in Logic (Mathematical logic and foundations)},
  VOLUME = {55},
  YEAR = {2015},
  MONTH = Jan,
  HAL_ID = {hal-01094195},
  HAL_VERSION = {v1},
}
